\documentclass[11pt]{article}

\usepackage{amsmath,amssymb,amsthm}
\usepackage{color}
\usepackage[noadjust]{cite}
\usepackage{graphicx}
\usepackage{epsfig}
\usepackage{breqn}
\usepackage[caption=false,font=footnotesize]{subfig}
\usepackage{array}
\usepackage{enumitem}
\usepackage{scalerel}

\newtheorem{theorem}{Theorem}[section]
\newtheorem{lemma}[theorem]{Lemma}

\newtheorem{prop}[theorem]{Proposition}

\allowdisplaybreaks
\newcommand\given[1][]{\:#1\vert\:}
\renewcommand\thesection{\arabic{section}}

\usepackage{relsize}

\newcommand\blfootnote[1]{%
  \begingroup
  \renewcommand\thefootnote{}\footnote{#1}%
  \addtocounter{footnote}{-1}%
  \endgroup
}

\usepackage{mathtools}
\makeatletter
\newcommand{\vast}{\bBigg@{4}}
\newcommand{\Vast}{\bBigg@{5}}

\makeatother

\allowdisplaybreaks
\renewcommand\thesection{\arabic{section}}
\newcommand{\bP}[1]{{\mathbb{P}}\left[{#1}\right]}

\newcommand{\fsquare}{\vrule height6pt width7pt depth1pt}   
\newcommand{\myproof}{{\hfill \\ \bf Proof. \ }}           
\newcommand{\myendpf}{\hfill\fsquare \\[0.1in]}             

\newcommand{\ER}{Erd\H{o}s-R\'enyi}
\newcommand{\EG}{Eschenauer-Gligor}
\newcommand{\crypto}{cryptographic}

\begin{document}

\title{Secure Connectivity of Heterogeneous Wireless Sensor Networks Under a Heterogeneous On-Off Channel Model}

\author{
Rashad~Eletreby and
        Osman~Ya\u{g}an \\
{\tt reletreby@cmu.edu}, {\tt oyagan@ece.cmu.edu} \\
Department of Electrical and Computer Engineering and CyLab\\
Carnegie Mellon University\\
Pittsburgh, PA 15213
}

\maketitle
\blfootnote{A preliminary version of some of the material was presented at the 54th Annual Allerton Conference on Communications, Control and Computing in 2016 \cite{EletrebyAllerton16} and at the IEEE International Symposium on Information Theory in 2017 \cite{EletrebyISIT17}. This work has been supported in part by the National Science Foundation through grant CCF-1617934. R. Eletreby was funded (in part) by the Dowd Fellowship from the College of Engineering at Carnegie Mellon University. The authors would like to thank Philip and Marsha Dowd for their financial support and encouragement.
}

\vspace{-10mm}

\begin{abstract}
Wireless sensor networks could be deployed in hostile environments where eavesdropping and node capture attacks are possible, inducing the need for \crypto{} protection. In this paper, we investigate the secure connectivity of wireless sensor networks utilizing the heterogeneous random key predistribution scheme, where each sensor node is classified as class-$i$ with probability $\mu_i$ for $i=1,\ldots,r$ with $\mu_i>0$ and $\sum_{i=1}^r \mu_i=1$. Before deployment, a class-$i$ sensor is given $K_i$ \crypto{} keys selected uniformly at random from a key pool of size $P$. 
After deployment, two sensor nodes can communicate securely over an available wireless channel if they share at least one \crypto{} key. In addition to the {\em shared-key connectivity} of the network as governed by the heterogeneous random key predistribution scheme, we consider the {\em wireless connectivity} of the network using a heterogeneous on-off channel model, where the channel between a class-$i$ node and a class-$j$ node is on (respectively, off) with probability $\alpha_{ij}$ (respectively, $1-\alpha_{ij}$) for $i,j=1,\ldots,r$ inducing a channel probability matrix $\pmb{\alpha}=\left[\alpha_{ij}\right]$. Collectively, two sensor nodes are adjacent if they i) share a \crypto{} key {\em and} ii) have a wireless channel in between that is on. 
We model the overall network using a composite random graph obtained by the intersection of inhomogeneous random key graphs $\mathbb{K}(n;\pmb{\mu},\pmb{K},P)$ with inhomogeneous \ER{} graphs $\mathbb{G}(n;\pmb{\mu}, \pmb{\alpha})$. The former graph is naturally induced by the heterogeneous random key predistribution scheme, while the latter is induced by the heterogeneous on-off channel model.  More specifically, two nodes are adjacent in the composite graph $\mathbb{K}(n;\pmb{\mu},\pmb{K},P) \cap \mathbb{G}(n;\pmb{\mu}, \pmb{\alpha})$ if they are i) adjacent in $\mathbb{K}(n;\pmb{\mu},\pmb{K},P)$, i.e., share a \crypto{} key and ii) adjacent in $\mathbb{G}(n;\pmb{\mu}, \pmb{\alpha})$, i.e., have an available wireless channel. Hence, edges in $\mathbb{K}(n;\pmb{\mu},\pmb{K},P) \cap \mathbb{G}(n;\pmb{\mu}, \pmb{\alpha})$ represent pairs of sensors which share a key and also have an available wireless channel in between. We investigate the connectivity of the composite random graph $\mathbb{K}(n;\pmb{\mu},\pmb{K},P) \cap \mathbb{G}(n;\pmb{\mu}, \pmb{\alpha})$ and present conditions (in the form of zero-one laws) on how to scale its parameters so that it i) has no secure node which is isolated and ii) is securely connected, both with high probability when the number of nodes gets large. We also present numerical results to support these zero-one laws in the finite-node regime.
\end{abstract}


\section{Introduction}

The proliferation of wireless sensor networks in multiple application domains, such as military applications, health care monitoring, among others, is attributed to their unique characteristics, such as their versatility, small-size, low-cost, ease of use, and scalability \cite{Akyildiz_2002, yick2008wireless, mainwaring2002wireless}. These features, however, give rise to unique security challenges that render wireless sensor networks vulnerable to a variety of security threats such as node capture attacks, node replication attacks, and eavesdropping \cite{security_survey}. Indeed, power-hungry cryptosystems such as {\em asymmetric} cryptosystems (public-key) are infeasible for securing large-scale wireless sensor networks that typically consist of battery-powered nodes with simple computation and communication architectures \cite{Gligor_2002,Haowen_2003, shi2004designing, liu2005establishing}. Accordingly, {\em symmetric} cryptosystems were shown to offer a faster and more energy-efficient alternative than their asymmetric counterpart, and they are deemed as the most feasible choice for securing wireless sensor networks \cite{Gligor_2002,Haowen_2003}. 

One key question associated with the use of symmetric cryptosystems is the design of key distribution mechanisms that facilitate the establishment of a secure communication infrastructure upon deploying the network and throughout its operation \cite{Haowen_2003, du2004key}. These mechanisms shall i) be fully distributed to avoid relying on any third party or a base station, ii) not assume any prior knowledge of post-deployment configuration, and iii) obey the hardware limitations of wireless sensor networks. Additionally, the resulting network shall be securely {\em connected} in a sense that there exists a secure communication path (not necessarily single-hop) between any pair of sensor nodes. The connectivity of the network is essential to its proper operation as it allows the exchange of control and data messages between any pair of sensor nodes.

Random key predistribution schemes were proposed in the seminal work of Eschenauer and Gligor \cite{Gligor_2002} to provide a feasible solution for key distribution in large-scale wireless sensor networks utilizing symmetric cryptosystems. In Eschenauer-Gligor scheme, each sensor node is assigned (before deployment) $K$ \crypto{} keys selected uniformly at random from a large key pool of size $P$. After deployment, two sensor nodes can communicate securely over an existing wireless channel if they share at least one key. The scheme does not require any prior knowledge of post deployment configuration and the communication infrastructure could be bootstrapped in a fully distributed manner. The connectivity of wireless sensor networks secured by Eschenauer and Gligor scheme was investigated in \cite{yagan2012zero, DiPietroTissec}, where scaling conditions for $K$ and $P$ were given to ensure that the resulting network is connected with high probability in the limit of large network size.

One inherent assumption with Eschenauer-Gligor scheme is that all sensor nodes are homogeneous, hence each node is given the same number $K$ of \crypto{} keys from the key pool. However, emerging wireless sensor networks are essentially complex and heterogeneous with different nodes performing different roles or equipped with different hardware capabilities \cite{Du2007_applications, Lu2008_applications, Wu2007_applications, Yarvis_2005}. Hence, different nodes could be assigned different number of keys depending on their roles or demands. For instance, a particular class of nodes may act as cluster heads which connect several clusters of nodes together. These cluster heads need to communicate with a large number of nodes in their vicinity and they are also expected to be more powerful than regular nodes. Thus, more keys should be given to the cluster heads to ensure high levels of connectivity and security.

To accommodate the emerging heterogeneity of wireless sensor networks, Ya\u{g}an proposed the {\em heterogeneous} random key predistribution scheme \cite{Yagan/Inhomogeneous} as a generalization of Eschenauer-Gligor scheme to account for the cases when the network comprises sensor nodes with varying level of resources and connectivity requirements. The scheme is characterized by $r$ different classes, where each node is classified as class-$i$ with probability $\mu_i$ with $\mu_i>0$ for $i=1,\ldots,r$ and $\sum_{i=1}^r \mu_i = 1$. A class-$i$ node is given $K_i$ \crypto{} keys selected uniformly at random (without replacement) from a large key pool of size $P$. Without loss of generality, it is assumed that $K_1 \leq K_2 \leq \ldots \leq K_r$. After deployment, two nodes can communicate securely over an existing channel if they share at least one key. The heterogeneous scheme gives rise to a class of random graphs known as {\em inhomogeneous} random key graphs $\mathbb{K}(n;\pmb{\mu},\pmb{K},P)$ \cite{Yagan/Inhomogeneous}, where each of the $n$ vertices is classified as class-$i$ with probability $\mu_i>0$ for $i=1,\ldots,r$ such that $\sum_{i=1}^r \mu_i=1$. A class-$i$ vertex $v_x$ is given a set $\Sigma_x$ of $K_i$ objects, selected uniformly at random (without replacement) from an object pool of size $P$. Two vertices $v_x$ and $v_y$ are adjacent if they share at least one object, i.e., if $\Sigma_x \cap \Sigma_y \neq \emptyset$. In \cite{Yagan/Inhomogeneous}, Ya\u{g}an derived scaling conditions for $\pmb{\mu}=\{\mu_1,\ldots,\mu_r\}$, $\pmb{K}$, and $P$ such that $\mathbb{K}(n;\pmb{\mu},\pmb{K},P)$ is connected with high probability in the limit of large network size. Essentially, the results given in \cite{Yagan/Inhomogeneous} provide guidelines on how to dimension the parameters of the heterogeneous random key predistribution scheme, i.e., $\pmb{\mu}$, $\pmb{K}$, and $P$, (with respect to the network size $n$) such that the resulting network is securely connected.

Note that edges in $\mathbb{K}(n;\pmb{\mu},\pmb{K},P)$ represent pairs of sensors that share at least one \crypto{} key, hence the model only encodes {\em shared-key} connectivity. In other words, it is assumed that all wireless channels are available and reliable, hence the only condition for two nodes to communicate securely is to share a \crypto{} key. In practice, the wireless channel is often unreliable and sensor nodes typically have limited communication ranges, hence, two sensor nodes which share a key may not eventually be adjacent due to the unavailability of their corresponding wireless channel. Accordingly, the secure connectivity of the network would not only be governed by the shared-key connectivity discussed above, but also by the wireless connectivity. As a result, the scaling conditions given in \cite{Yagan/Inhomogeneous} would be too optimistic for real-world deployments characterized by unreliable wireless media.

In this paper, we investigate the connectivity of wireless sensor networks secured by the heterogeneous random key predistribution scheme under a {\em heterogeneous on-off} channel model. In this channel model, the wireless channel between a class-$i$ node and a class-$j$ node is on with probability $\alpha_{ij}$ and off with probability $1-\alpha_{ij}$, independently. This gives rise to a $r \times r$ channel probability matrix $\pmb{\alpha}$ where the element at the $i$th row and $j$th column is given by $\alpha_{ij}$. The heterogeneous on-off channel model accounts for the fact that different nodes could have different radio capabilities, or could be deployed in locations with different channel characteristics. In addition, it offers the flexibility of modeling several interesting scenarios, such as when nodes of the same type are more (or less) likely to be adjacent with one another than with nodes belonging to other classes. The heterogeneous on-off channel model gives rise to inhomogeneous Erd\H{o}s-R\'enyi graphs \cite{devroye2014connectivity,bollobas2007phase}, denoted hereafter by $\mathbb{G}(n,\pmb{\mu}, \pmb{\alpha})$. In these graphs, each of the $n$ vertices is classified as class-$i$ with probability $\mu_i>0$ such that $\sum_{i=1}^r \mu_i=1$. Two vertices $v_x$ and $v_y$, which belong to class-$i$ and class-$j$, respectively, are adjacent if $B(\alpha_{ij})=1$, where $B(\alpha_{ij})$ denotes a Bernoulli random variable with success probability $\alpha_{ij}$. 

Edges in inhomogeneous random keys graphs encode shared-key relationships, while edges in inhomogeneous \ER{} graphs encode the availability of wireless channels. Hence, the overall network can be modeled by a {\em composite} random graph model formed by the {\em intersection} of an inhomogeneous random key graph with an inhomogeneous \ER{} graph. We denote the intersection graph $\mathbb{K}(n;\pmb{\mu},\pmb{K},P) \cap \mathbb{G}(n;\pmb{\mu},\pmb{\alpha})$ by $\mathbb{H}(n;\pmb{\mu},\pmb{K},P,\pmb{\alpha})$. An edge exists in $\mathbb{H}(n;\pmb{\mu},\pmb{K},P,\pmb{\alpha})$ only if it exists in $\mathbb{K}(n;\pmb{\mu},\pmb{K},P)$, i.e., both nodes share a key, and $\mathbb{G}(n;\pmb{\mu},\pmb{\alpha})$, i.e., both nodes share a wireless channel. Hence, edges in $\mathbb{H}(n;\pmb{\mu},\pmb{K},P,\pmb{\alpha})$ represent pairs of sensors that both i) share a key and ii) have a wireless channel in between that is on.

We investigate the connectivity of the composite random graph $\mathbb{H}(n;\pmb{\mu},\pmb{K},P,\pmb{\alpha})$ and present conditions (in the form of zero-one laws) on how to scale its parameters, i.e., $\pmb{\mu}$, $\pmb{K}$, $P$, and $\pmb{\alpha}$, so that it i) has no secure node which is isolated and ii) is securely connected, both with high probability when the number of nodes gets large. Essentially, our results provide design guidelines on how to choose the parameters of the heterogeneous random key predistribution scheme such that the resulting wireless sensor network is securely connected under a heterogeneous on-off channel model. Our results are supported by a simulation study demonstrating that despite their asymptotic nature, our results can in fact be useful in designing finite-node wireless sensor network so that they achieve secure connectivity with high probability.

We close with a word on notation and conventions in use. All limiting statements, including asymptotic equivalence are considered with the number of sensor nodes $n$ going to infinity. The random variables (rvs) under consideration are all defined on the same probability triple $(\Omega,\mathcal{F},\mathbb{P})$. Probabilistic statements are made with respect to this probability measure $\mathbb{P}$, and we denote the corresponding expectation by $\mathbb{E}$. The indicator function of an event $E$ is denoted by $\pmb{1}[E]$. We say that an event holds with high probability (whp) if it holds with probability $1$ as $n \rightarrow \infty$. For any discrete set $S$, we write $|S|$ for its cardinality.
In comparing
the asymptotic behaviors of the sequences $\{a_n\},\{b_n\}$,
we use
$a_n = o(b_n)$,  $a_n=\omega(b_n)$, $a_n = O(b_n)$, $a_n = \Omega(b_n)$, and
$a_n = \Theta(b_n)$, with their meaning in
the standard Landau notation. We also use $a_n \sim b_n$ to denote the asymptotic equivalence $\lim_{n \to \infty} {a_n}/{b_n}=1$.

\section{The Model}
\subsection{Shared-key connectivity: Inhomogeneous random key graphs $\mathbb{K}(n;\pmb{\mu},\pmb{K},P)$}
Consider $n$ sensor nodes labeled as $v_1, v_2, \ldots,v_n$, where each node is classified into one of $r$ classes with a probability distribution $\pmb{\mu}=\{\mu_1,\mu_2,\ldots,\mu_r\}$ with $\mu_i >0$ for $i=1,\ldots,r$ and $\sum_{i=1}^r \mu_i=1$. A class-$i$ node is assigned $K_i$ cryptographic keys selected uniformly at random (without replacement) from a key pool of size $P$. It follows that the key ring $\Sigma_x$ of node $x$ is a $\mathcal{P}_{K_{t_x}}$-valued random variable (rv) where $\mathcal{P}_{K_{t_x}}$ denotes the collection of all subsets of $\{1,\ldots,P\}$ with exactly $K_{t_x}$ elements and $t_x$ denotes the class of node $v_x$. The rvs $\Sigma_1, \Sigma_2, \ldots, \Sigma_n$ are then i.i.d. with
\begin{equation} \nonumber
\mathbb{P}[\Sigma_x=S \mid t_x=i]= \dbinom P{K_i}^{-1}, \quad S \in \mathcal{P}_{K_i}.
\nonumber
\end{equation}
Let $\pmb{K}=\{K_1,K_2,\ldots,K_r\}$ and assume without loss of generality that $K_1 \leq K_2 \leq \ldots \leq K_r$. Consider a random graph $\mathbb{K}$ induced on the vertex set $\mathcal{V}=\{v_1,\ldots,v_n\}$ such that a pair of distinct nodes $v_x$ and $v_y$ are adjacent in $\mathbb{K}$, denoted by $v_x \sim_K v_y$, if they have at least one key in common, i.e.,
\begin{equation}
v_x \sim_K v_y \quad \text{if} \quad \Sigma_x \cap \Sigma_y \neq \emptyset.
\label{adjacency_condition}
\end{equation}

The adjacency condition (\ref{adjacency_condition}) defines inhomogeneous random key graphs denoted by $\mathbb{K}(n;\pmb{\mu},\pmb{K},P)$  \cite{Yagan/Inhomogeneous}. This model is also known in the literature  as
the {\em general random intersection graph}; e.g., see \cite{Zhao_2014,Rybarczyk,Godehardt_2003}. 
 The probability $p_{ij}$ that a class-$i$ node and a class-$j$ node are adjacent is given by
\begin{equation}
p_{ij}=1-\frac{\binom {P-K_i}{K_j}}{\binom {P}{K_j}}
\label{eq:osy_edge_prob_type_ij}
\end{equation}
as long as $K_i + K_j \leq P$; otherwise if $K_i +K_j > P$, we  have $p_{ij}=1$.
Let $\lambda_i$ denote the \textit{mean} probability that a class-$i$ node is connected to another node in $\mathbb{K}(n;\pmb{\mu},\pmb{K},P)$. We have
\begin{align}
\lambda_i & =\sum_{j=1}^r \mu_j p_{ij}.
 \label{eq:osy_mean_edge_prob_in_RKG}
\end{align}

\subsection{Wireless connectivity: Inhomogeneous Erd\H{o}s-R\'enyi graphs $\mathbb{G}(n;\pmb{\mu},\pmb{\alpha})$}

In practical deployments of wireless sensor networks, nodes typically have limited communication ranges and the channel between two nodes may not be available, e.g., due to excessive interference. In other words, two sensor nodes which share a key may not eventually be adjacent due to the unavailability of their corresponding wireless channel. Hence, the secure connectivity of the network would not only be governed by the shared-key connectivity discussed above, but also by the wireless connectivity.

In modeling the wireless connectivity of the network, we utilize a heterogeneous on-off channel model, where the wireless channel between a class-$i$ node and a class-$j$ node is on (respectively, off) with probability $\alpha_{ij}$ (respectively, $1-\alpha_{ij}$) for $i,j=1,\ldots,r$. Note that the heterogeneous on-off channel model accounts for the fact that different nodes could have different radio capabilities, or could be deployed in locations with different channel characteristics. This is indeed a generalization of the uniform on-off channel model, where the channel between any two nodes is on (respectively, off) with probability $\alpha$ (respectively, $1-\alpha$) regardless of the corresponding classes. Hence, the heterogeneous on-off channel model offers the flexibility of modeling several interesting scenarios, such as when nodes of the same type are more (or less) likely to be adjacent with one another than with nodes belonging to other classes.

Consider a random graph $\mathbb{G}$ induced on the vertex set $\mathcal{V}=\{v_1,\ldots,v_n\}$ such that each node is classified into one of the $r$ classes with a probability distribution $\pmb{\mu}=\{\mu_1,\mu_2,\ldots,\mu_r\}$ with $\mu_i >0$ for $i=1,\ldots,r$ and $\sum_{i=1}^r \mu_i=1$. Then, a distinct class-$i$ node $v_x$ and a distinct class-$j$ node $v_y$ are adjacent in $\mathbb{G}$, denoted by $v_x \sim_G v_y$, if $B_{xy}(\alpha_{ij})=1$ where $B_{xy}(\alpha_{ij})$ denotes a Bernoulli rv with success probability $\alpha_{ij}$. This gives rise to an $ r \times r$ edge probability matrix $\pmb{\alpha}$ where $\alpha_{ij}$ denotes the element of row $i$ and column $j$ of $\pmb{\alpha}$. The aforementioned adjacency conditions induces the inhomogeneous Erd\H{o}s-R\'enyi graph $\mathbb{G}(n;\pmb{\mu},\pmb{\alpha})$ on the vertex set $\mathcal{V}$, which has received interest recently \cite{devroye2014connectivity,bollobas2007phase}.

Although the on-off channel model may be considered too simple, it allows a comprehensive analysis of the properties of interest and is often a good approximation of more realistic channel models, e.g., the disk model \cite{Gupta99}. In fact, the simulations results in \cite{EletrebyKConn} suggest that the $k$-connectivity behavior of wireless sensor networks secured by the heterogeneous random key predistribution scheme under the uniform on-off channel model (where $\alpha_{ij}=\alpha$ for $i,j=1,\ldots,r$) is asymptotically equivalent to that under the more-realistic disk model.


\subsection{The composite random graph $\pmb{H}(n;\pmb{\mu},\pmb{K},P,\pmb{\alpha}):=\mathbb{K}(n;\pmb{\mu},\pmb{K},P) \cap \mathbb{G}(n;\pmb{\mu},\pmb{\alpha})$}

Each of the above two random graph models captures a particular notion of connectivity, namely shared-key connectivity and wireless connectivity, respectively. In what follows, we construct a random graph model that jointly considers both notions, hence, it accurately describes practical deployments of wireless sensor networks, where two nodes are adjacent if they both share a key {\em and} have an available wireless channel in between.

We consider a composite random graph obtained by the intersection of inhomogeneous random key graphs $\mathbb{K}(n;\pmb{\mu},\pmb{K},P)$ with inhomogeneous \ER{} graphs $\mathbb{G}(n;\pmb{\mu},\pmb{\alpha})$. We denote the intersection graph by $\mathbb{H}(n;\pmb{\mu},\pmb{K},P,\pmb{\alpha})$, i.e., $\mathbb{H}(n;\pmb{\mu},\pmb{K},P,\pmb{\alpha}):=\mathbb{K}(n;\pmb{\mu},\pmb{K},P) \cap \mathbb{G}(n;\pmb{\mu},\pmb{\alpha})$. Hence, edges in the intersection graph $\mathbb{H}(n;\pmb{\mu},\pmb{K},P,\pmb{\alpha})$ represent pairs of sensor which i) share a key and ii) have a wireless channel in between that is on. In particular, a distinct class-$i$ node $v_x$ is adjacent to a distinct class-$j$ node $v_y$ in $\mathbb{H}$ if and only if they are adjacent in both $\mathbb{K}$ \textit{and} $\mathbb{G}$. 

To simplify the notation, we let $\pmb{\theta}=(\pmb{K},P)$, and $\pmb{\Theta}=(\pmb{\theta},\pmb{\alpha})$. By independence, we see that the probability of edge assignment  between a class-$i$ node $v_x$ and a class-$j$ node $v_y$ in $\mathbb{H}(n;\pmb{\mu},\pmb{\Theta})$ is given by
\begin{equation} \nonumber
\mathbb{P}[v_x \sim v_y \mid t_x=i,t_y=j]=\alpha_{ij} p_{ij}
\end{equation}
Similar to (\ref{eq:osy_mean_edge_prob_in_RKG}), we denote the mean edge probability for a class-$i$ node in $\mathbb{H}(n;\pmb{\mu},\pmb{\Theta})$ as $\Lambda_i$. It is clear that
\begin{align} 
\Lambda_i = \sum_{j=1}^r \mu_j \alpha_{ij} p_{ij}, \quad i=1,\ldots, r.
\label{eq:osy_mean_edge_prob_in_system}
\end{align}

We write $\Lambda_m$ to denote the {\em minimum} mean edge probability in $\mathbb{H}(n;\pmb{\mu},\pmb{\Theta})$, i.e.,  
\begin{equation}
m:=\arg \min_i \Lambda_i.
\label{eq:min_mean_degree}
\end{equation}
We further let $\alpha_{\mathrm{min}} := \min_{i,j} \{ \alpha_{ij} \}$ and $\alpha_{\mathrm{max}} := \max_{i,j} \{\alpha_{ij}\}$.  Finally, we define $d$ and $s$ as follows
\begin{align}
&d:=\arg \max_j \{\alpha_{mj} \}, \label{eq:HetER0}\\
&s:=\arg \max_j \{\alpha_{mj} p_{mj} \}. \label{eq:HetER_s}
\end{align}

Throughout, we assume that the number of classes $r$ is fixed and does not scale with $n$, and so are the probabilities $\mu_1, \ldots,\mu_r$. All of the remaining parameters are assumed to be scaled with $n$.

\section{Main Results and Discussion}
We refer to a mapping $K_1,\ldots,K_r,P: \mathbb{N}_0 \rightarrow \mathbb{N}_0^{r+1}$ as a \textit{scaling} (for inhomogeneous random key graphs) if\begin{equation}
1 \leq K_{1,n} \leq K_{2,n} \leq \ldots \leq K_{r,n} \leq P_n/2
\label{scaling_condition_K}
\end{equation}
hold  for all $n=2,3,\ldots$. Similarly any mapping $\pmb{\alpha}=\{ \alpha_{ij} \}: \mathbb{N}_0 \rightarrow (0,1)^{r \times r}$ defines a scaling for inhomogeneous \ER{} graphs. A mapping $\pmb{\Theta} : \mathbb{N}_0 \rightarrow \mathbb{N}_0^{r+1} \times (0,1)^{r \times r}$ defines a scaling for the intersection graph $\mathbb{H}(n;\pmb{\mu},\pmb{\Theta})$ given that condition (\ref{scaling_condition_K}) holds. We remark that under (\ref{scaling_condition_K}), the edge probabilities $p_{ij}$ will be given by
(\ref{eq:osy_edge_prob_type_ij}).

\subsection{Results}

We first present a zero-one law for the absence of isolated nodes in 
$\mathbb{H}(n;\pmb{\mu},\pmb{\Theta}_n)$.
\begin{theorem}
\label{theorem:isolated_nodes}
\sl
Consider a probability distribution $\pmb{\mu}=\{\mu_1,\mu_2,\ldots,\mu_r\}$ with $\mu_i >0$ for $i=1,\ldots,r$, a scaling $K_1,\ldots,K_r,P: \mathbb{N}_0 \rightarrow \mathbb{N}_0^{r+1}$, and a scaling $\pmb{\alpha}=\{\alpha_{ij}\}: \mathbb{N}_0 \rightarrow (0,1)^{r \times r}$ such that
\begin{equation}
\Lambda_m(n) \sim c \frac{\log n}{n}
\label{scaling_condition_KG}
\end{equation}
holds for some $c>0$.

i) If
\begin{equation*}
\lim_{n \to \infty} \alpha_{md}(n) \log n=0 \qquad \text{or} \qquad  \lim_{n \to \infty} \alpha_{mm}(n) \log n=\alpha^{*} \in (0,\infty]
\end{equation*}
holds, then we have
\begin{equation} \nonumber
\lim_{n\to\infty} \mathbb{P} \left[  \mathbb{H}(n;\pmb{\mu},\pmb{\Theta}_n) \text{ has no isolated nodes}  \right]= 0 \qquad \text{ if } c<1
\end{equation}

ii) We have
\begin{equation} \nonumber
\lim_{n\to\infty} \mathbb{P} \left[  \mathbb{H}(n;\pmb{\mu},\pmb{\Theta}_n) \text{ has no isolated nodes}  \right]= 1 \qquad \text{ if } c>1
\end{equation}
\end{theorem}

Next, we present an analogous result for connectivity.
\begin{theorem}
\label{theorem:connectivity}
\sl
Consider a probability distribution $\pmb{\mu}=\{\mu_1,\mu_2,\ldots,\mu_r\}$ with $\mu_i >0$ for $i=1,\ldots,r$, a scaling $K_1,\ldots,K_r,P: \mathbb{N}_0 \rightarrow \mathbb{N}_0^{r+1}$, and a scaling $\pmb{\alpha}=\{\alpha_{ij}\}: \mathbb{N}_0 \rightarrow (0,1)^{r \times r}$ such that (\ref{scaling_condition_KG}) holds for some $c>0$.

i) If
\begin{equation*}
\lim_{n \to \infty} \alpha_{md}(n) \log n=0 \qquad \text{or} \qquad  \lim_{n \to \infty} \alpha_{mm}(n) \log n=\alpha^{*} \in (0,\infty]
\end{equation*}
holds, then we have
\begin{equation} \nonumber
\lim_{n\to\infty} \mathbb{P} \left[\mathbb{H}(n;\pmb{\mu},\pmb{\Theta}_n) \text{ is connected}  \right]= 0 \quad \text{ if } c<1
\end{equation}

ii) If
\begin{equation}
P_n \geq \sigma n, \quad n=1, 2, \ldots
\label{eq:conn_Pn2}
\end{equation}
for some $\sigma>0$, and
\begin{equation}
\alpha_{\min}(n) p_{1r}(n) = \Omega \left( \frac{\log n}{n} \right)
\label{eq:conn_K1r}
\end{equation}

\begin{equation}
\frac{K_{r,n}}{K_{1,n}} = o \left( \log n \right)
\label{eq:bounding_variance_of_K}
\end{equation}

\begin{equation}
\frac{\alpha_{\max}(n)}{\alpha_{\min}(n)} = O \left( \left( \log n \right)^\tau \right)
\label{eq:bounding_variance_of_alpha}
\end{equation}
for any finite $\tau>0$.
Then, we have
\begin{equation} \nonumber
\lim_{n\to\infty}  \mathbb{P} \left[\mathbb{H}(n;\pmb{\mu},\pmb{\Theta}_n) \text{ is connected}  \right]= 1 \quad \text{ if } c>1
\end{equation}
\end{theorem}
The scaling condition (\ref{scaling_condition_KG}) will often be used in the form
\begin{equation} \label{scaling_condition_KG_v2}
\Lambda_m(n)=c_n \frac{\log n}{n}, \ n=2,3,\ldots
\end{equation}
with $\lim_{n\to\infty} c_n=c>0$. Also, condition (\ref{eq:conn_K1r}) will often be used in the form
\begin{equation}
\alpha_{\min}(n) p_{1r}(n) \geq \rho \frac{\log n}{n}, \quad \text{for } \rho>0 \text{ and } n=2,3,\ldots
\label{eq:conn_K1r_modified}
\end{equation}

\subsection{Discussion}
Theorems~\ref{theorem:isolated_nodes} and \ref{theorem:connectivity} state that $\mathbb{H}(n;\pmb{\mu},\pmb{\Theta}_n)$ has no isolated node (and is connected) with high probability if the minimum mean degree, i.e., $n \Lambda_m$, is scaled as $(1+\epsilon) \log n$ for some $\epsilon > 0$. On the other hand, if this minimum mean degree scales as $(1-\epsilon) \log n$ for some $\epsilon > 0$, then with high probability $\mathbb{H}(n;\pmb{\mu},\pmb{\Theta}_n)$ has an isolated node, and hence is not connected. The resemblance of the results presented in Theorem~\ref{theorem:isolated_nodes} and Theorem~\ref{theorem:connectivity} indicates that  absence of isolated nodes and connectivity are asymptotically equivalent properties for $\mathbb{H}(n;\pmb{\mu},\pmb{\Theta}_n)$. Similar observations were made
for other well-known random graph models as well; e.g., inhomogeneous random key graphs \cite{Yagan/Inhomogeneous}, \ER{} graphs \cite{Bollobas}, and (homogeneous) random key graphs \cite{yagan2012zero}.

Note that if the matrix $\pmb{\alpha}$ is designed in such a way that $\alpha_{ii} = \max_j \{\alpha_{ij}\}$, i.e., two nodes of the same type are more likely to be adjacent in $\mathbb{G}(n;\pmb{\mu}, \pmb{\alpha})$, then we have $\alpha_{md} = \alpha_{mm}$ and the condition of the zero-law of Theorems~\ref{theorem:isolated_nodes} and \ref{theorem:connectivity} would collapse to i) $\lim_{n \to \infty} \alpha_{mm}(n) \log n = 0 $ or ii) $\lim_{n \to \infty} \alpha_{mm}(n) \log n \in (0, \infty]$. At this point, the zero-law follows even when the sequence $\alpha_{mm} \log n$ does not have a limit by virtue of the {\em subsubsequence principle} \cite[p. 12]{JansonLuczakRucinski} (see also \cite[Section~7.3]{EletrebyKConn}). In other words, if $\alpha_{md} = \alpha_{mm}$, then the zero-law of Theorems~\ref{theorem:isolated_nodes} and \ref{theorem:connectivity} follows without any conditions on the sequence $\alpha_{mm}(n) \log n $. 

We now comment on the additional technical conditions needed for the one-law of Theorem~\ref{theorem:connectivity}. Condition (\ref{eq:conn_Pn2}) is likely to be needed in practical deployments of wireless sensor networks in order to ensure the {\em resilience} of the network against node capture attacks; e.g., see \cite{Gligor_2002,DiPietroTissec}. To see this, assume that an adversary captures a number of sensors, compromising all the keys that belong to the captured nodes. If $P_n = o(n)$, contrary to  (\ref{eq:conn_Pn2}), then it would be possible for the adversary to compromise $\Omega(P_n)$ keys by capturing only  $o(n)$ sensors (whose type does not matter). In this case, the wireless sensor network would fail to exhibit the {\em unassailability} property \cite{MeiPanconesiRadhakrishnan2008,YM_ToN} and would be deemed as vulnerable against adversarial attacks.
We remark that (\ref{eq:conn_Pn2}) was required in \cite{Rashad/Inhomo, Jun/K-Connectivity, EletrebyKConn, Yagan/Inhomogeneous} in similar settings to ours.

Condition (\ref{eq:conn_K1r}) provides a non-trivial lower bound on the edge probability $\alpha_{\min}(n) p_{1r}(n)$ and is enforced mainly for technical reasons for the proof of the one-law of Theorem~\ref{theorem:connectivity} to work. Note that it is easy to show that $\alpha_{\min}(n) p_{1r}(n) = O\left( \log n / n \right)$ from (\ref{scaling_condition_KG_v2}) (see Lemma~\ref{lemma:isolated_1} for a proof), however, the scaling condition given by (\ref{scaling_condition_KG_v2}) does not provide any non-trivial lower-bound on the product $\alpha_{\min}(n) p_{1r}(n)$. Observe that, even with condition (\ref{eq:conn_K1r}), our results do not require {\em each} edge probability to scale as $\log n / n $, in contrast to the results given in \cite{devroye2014connectivity} on the connectivity of inhomogeneous \ER{} graphs. In particular, the probability of an edge between a class-$i$ node and a class-$j$ node was set to $\kappa \left(i,j \right) \log n / n$ in \cite{devroye2014connectivity}, where $\kappa \left(i,j \right)$ returns a positive real number for each pair $(i,j)$; i.e., each individual edge was scaled as $ \Theta(\log n / n)$.

Condition (\ref{eq:bounding_variance_of_K}) is also enforced mainly for technical reasons and it takes away from the flexibility of assigning very small key key rings to a certain fraction of sensors when connectivity is considered. An equivalent condition was also needed in \cite{Yagan/Inhomogeneous} for establishing the one-law for connectivity in inhomogeneous random key graphs. We refer the reader to \cite[Section 3.2]{Yagan/Inhomogeneous} for an extended discussion on the feasibility of (\ref{eq:bounding_variance_of_K}) for real-world implementations of wireless sensor networks. Condition (\ref{eq:bounding_variance_of_alpha}) also limits the flexibility of assigning very small values for $\alpha_{\min}$, but it is much milder than condition (\ref{eq:bounding_variance_of_K}) in a sense that it requires $\alpha_{\max} (n)/ \alpha_{\min}(n)$ to be $O\left( \left( \log n \right)^\tau \right)$ for some finite $\tau>0$, i.e., one can still afford to have a large deviation between $\alpha_{\min}(n)$ and $\alpha_{\max}(n)$ as compared to the case if $\alpha_{\max} (n)/ \alpha_{\min}(n)$ had to be scaled as $o(\log n)$, similar to the case in (\ref{eq:bounding_variance_of_K}).

We close by providing a concrete example that demonstrates how all the conditions required by Theorem \ref{theorem:connectivity}
can be met in a real-world implementation. Consider a sensor network consisting of two classes, i.e., $r=2$. Pick any probability distribution $\pmb{\mu}=\{\mu_1, \mu_2\}$ with $\mu_i > 0$ for all $i=1,2$. Set $P_n = \left \lceil  n \log n \right \rceil $ as well as 
\[
K_{1,n} =\left \lceil \frac{(\log n)^{1/2+\varepsilon}}{\sqrt{\alpha_{\min}(n)}}  \right \rceil  \quad \textrm{and} \quad K_{2,n} = \left \lceil \frac{(1 + \varepsilon)(\log n)^{3/2-\varepsilon}}{\mu_2 \sqrt{\alpha_{\min}(n)}} \right \rceil
\]
with any $0 < \varepsilon < 0.5$. Observe that the above selection satisfies (\ref{eq:conn_Pn2}) as well as (\ref{eq:bounding_variance_of_K}). Next, set
\begin{equation}\nonumber
\pmb{\alpha} = 
\alpha_{\min}(n)
\left[
\begin{matrix}
\frac{1+\epsilon}{\mu_1} \left(\log n \right)^{1-2 \epsilon} & 1 \\
1 & \frac{\mu_2}{1+\epsilon} \left( \log n\right)^{1+2 \epsilon}
\end{matrix}
\right]
\end{equation}
Note that the above selection satisfies (\ref{eq:bounding_variance_of_alpha}) with $\tau=1+2\epsilon$. For simplicity, assume that $\lambda_1(n)=o(1)$ which implies that $p_{1j}(n)=o(1)$ for $j=1,2$. In this case, we have $p_{1j}(n) \sim \frac{K_{1,n} K_{j,n}}{P_n}$ for $j=1,2$ (see \cite[Lemma 4.2]{Yagan/Inhomogeneous}). With this parameter selection, we have
\begin{equation}\nonumber
\alpha_{\min}(n) p_{12}(n) \sim \alpha_{\min}(n)  \frac{K_{1,n} K_{2,n}}{P_n} = \frac{1+\epsilon}{\mu_2} \frac{\log n}{n}
\end{equation}
which satisfies (\ref{eq:conn_K1r}).

Finally, observe that with the above parameter selection, both $\Lambda_1(n)$ and $\Lambda_2(n)$ are strictly larger than $\log n / n $. Hence, in view of Theorem \ref{theorem:connectivity}, the resulting network will be connected with high probability. Of course, there are many other parameter scalings that one can choose.

\subsection{Comparison with related work}
The connectivity (respectively, $k$-connectivity)  of wireless sensor networks secured by the classical \EG{} scheme under a {\em uniform} on/off channel model was investigated in \cite{Yagan/EG_intersecting_ER} (respectively, \cite{Jun/K-Connectivity}). The network was modeled by a composite random graph formed by the intersection of random key graphs $\mathbb{K}(n;K,P)$ (induced by \EG{} scheme) with \ER{} graphs $\mathbb{G}(n;\alpha)$ (induced by the uniform on-off channel model). Our paper generalizes this model to heterogeneous setting where different nodes could be given different number of keys depending on their respective classes and the availability of a wireless channel between two nodes depends on their respective classes. Hence, our model highly resembles emerging wireless sensor networks which are essentially complex and heterogeneous.

In \cite{Yagan/Inhomogeneous}, Ya\u{g}an considered the connectivity of wireless sensor networks secured by the heterogeneous random key predistribution scheme under the full visibility assumption, i.e., all wireless channels are available and reliable, hence the only condition for two nodes to be adjacent is to share a key. It is clear that the full visibility assumption is not likely to hold in most practical deployments of wireless sensor networks as the wireless medium is typically unreliable. Our paper extends the results given in \cite{Yagan/Inhomogeneous} to more practical scenarios where the wireless connectivity is taken into account through the heterogeneous on-off channel model. In fact, by setting  
$\alpha_{ij}(n)=1$ for $i,j=1,\ldots,r$ and each $n=1,2,\ldots$ (i.e., by assuming that all wireless channels are {\em on}), our results reduce to those given in
\cite{Yagan/Inhomogeneous}. 

In comparison with the existing literature on similar models, our result can be seen to extend the work by Eletreby and Ya\u{g}an in \cite{Rashad/Inhomo} (respectively, \cite{EletrebyKConn}). Therein, the authors established a zero-one law for the $1$-connectivity (respectively, $k$-connectivity) of 
$\mathbb{K} (n;\pmb{\mu},\pmb{K},P) \cap \mathbb{G}(n;\alpha)$, i.e., for a wireless sensor network under the {\em heterogeneous} key predistribution scheme and a {\em uniform} on-off channel model. Although these results form a crucial starting point towards the analysis of the heterogeneous key predistribution scheme under a wireless connectivity model, they are limited to uniform on-off channel model where all channels are on (respectively, off) with the same probability $\alpha$ (respectively, $1-\alpha$). The heterogeneous on-off channel model accounts for the fact that different nodes could have different radio capabilities, or could be deployed in locations with different channel characteristics. In addition, it offers the flexibility of modeling several interesting scenarios, such as when nodes of the same type are more (or less) likely to be adjacent with one another than with nodes belonging to other classes. Indeed, by setting $\alpha_{ij}(n) = \alpha$ for $i,j=1,\ldots,r$ and each $n=1,2,\ldots$, our results reduce to those given in \cite{Rashad/Inhomo}.

\section{Numerical Results}
\label{sec:numerical}
In this section, we present a simulation study to validate our results in the finite-node regime. In all experiments, we fix the number of nodes at $n = 500$, the size of the key pool at $P = 10^4$, and the number of experiments to $400$.

\begin{figure}[t]
\centerline{\includegraphics[scale=0.6]{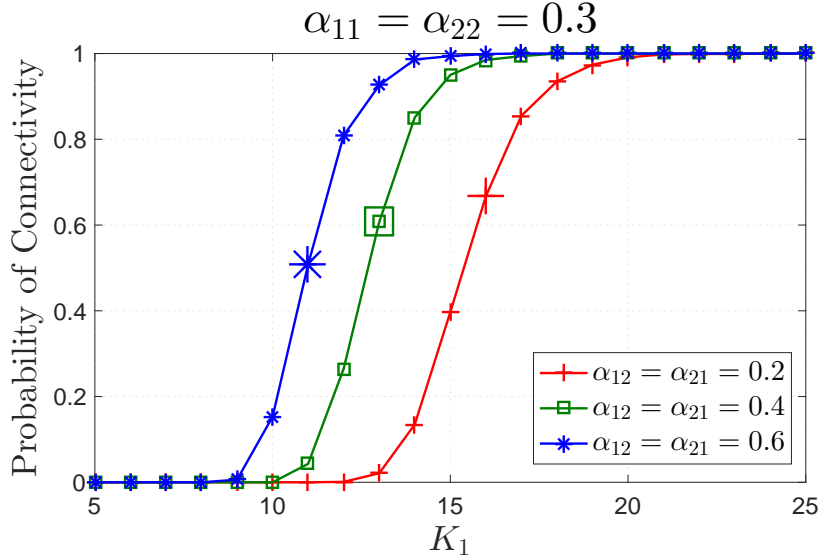}}
\caption{Empirical probability that $\mathbb{H}(n;\pmb{\mu},\pmb{\Theta})$ is connected as a function of $\pmb{K}$ for $\alpha_{12} = 0.2$, $\alpha_{12} = 0.4$, and $\alpha_{12} = 0.6$. We set $\alpha_{11}=\alpha_{22}=0.3$. Highlighted symbols stand for the critical threshold of connectivity asserted by Theorem~\ref{theorem:connectivity}.}
\label{fig:1}
\end{figure}

In Figure~\ref{fig:1}, we set the channel matrix to
\begin{equation*}
\pmb{\alpha}=
  \begin{bmatrix}
    0.3 & \alpha_{12} \\
    \alpha_{12} & 0.3
  \end{bmatrix}
\end{equation*}
and consider three different values for the parameter $\alpha_{12}$, namely, $\alpha_{12} = 0.2$, $\alpha_{12} = 0.4$, and $\alpha_{12} = 0.6$. We also vary $K_1$ (i.e., the smallest key ring size) from $5$ to $25$. The number of classes is fixed to $2$, with $\pmb{\mu}=\{0.5,0.5\}$. For each value of $K_1$, we set $K_2=K_1+5$. For each parameter pair $(\pmb{K}, \pmb{\alpha})$, we generate $400$ independent samples of the graph $\mathbb{H}(n;\pmb{\mu},\pmb{\Theta})$ and count the number of times (out of a possible $400$) that the obtained graphs i) have no isolated nodes and ii) are connected. Dividing the counts by $400$, we obtain the (empirical) probabilities for the events of interest.  In all cases considered here, we observe that $\mathbb{H}(n;\pmb{\mu},\pmb{\Theta})$ is connected whenever it has no isolated nodes yielding the same empirical probability for both events. This confirms the asymptotic equivalence of the connectivity and absence of isolated nodes properties in $\mathbb{H}(n;\pmb{\mu},\pmb{\Theta}_n)$ as is illustrated in Theorems~\ref{theorem:isolated_nodes} and ~\ref{theorem:connectivity}.

For each value of $\alpha_{12}$, we show the critical threshold of connectivity given by Theorem~\ref{theorem:connectivity} in the form of highlighted symbols. More specifically, highlighted symbols stand for the minimum integer value of $K_1$ that satisfies
\begin{equation}
\Lambda_m(n)=\sum_{j=1}^2 \mu_j \alpha_{mj} \left( 1- \frac{\binom{P-K_j}{K_m}}{\binom{P}{K_m}} \right) >\frac{\log n}{n}.
\label{eq:numerical_critical}
\end{equation}
upon noting that $K_2=  K_1+5$. We see from Figure~\ref{fig:1} that the probability of connectivity transitions from zero to one within relatively small variations of $K_1$. Moreover, the critical values of $K_1$ obtained by (\ref{eq:numerical_critical}) lie within this transition interval and correspond to high probability of connectivity. Note that for each parameter pair $(\pmb{K},\pmb{\alpha})$ in Figure~\ref{fig:1}, we have $\Lambda_m=\Lambda_1$ by construction.

\begin{figure}[t]
\centerline{\includegraphics[scale=0.6]{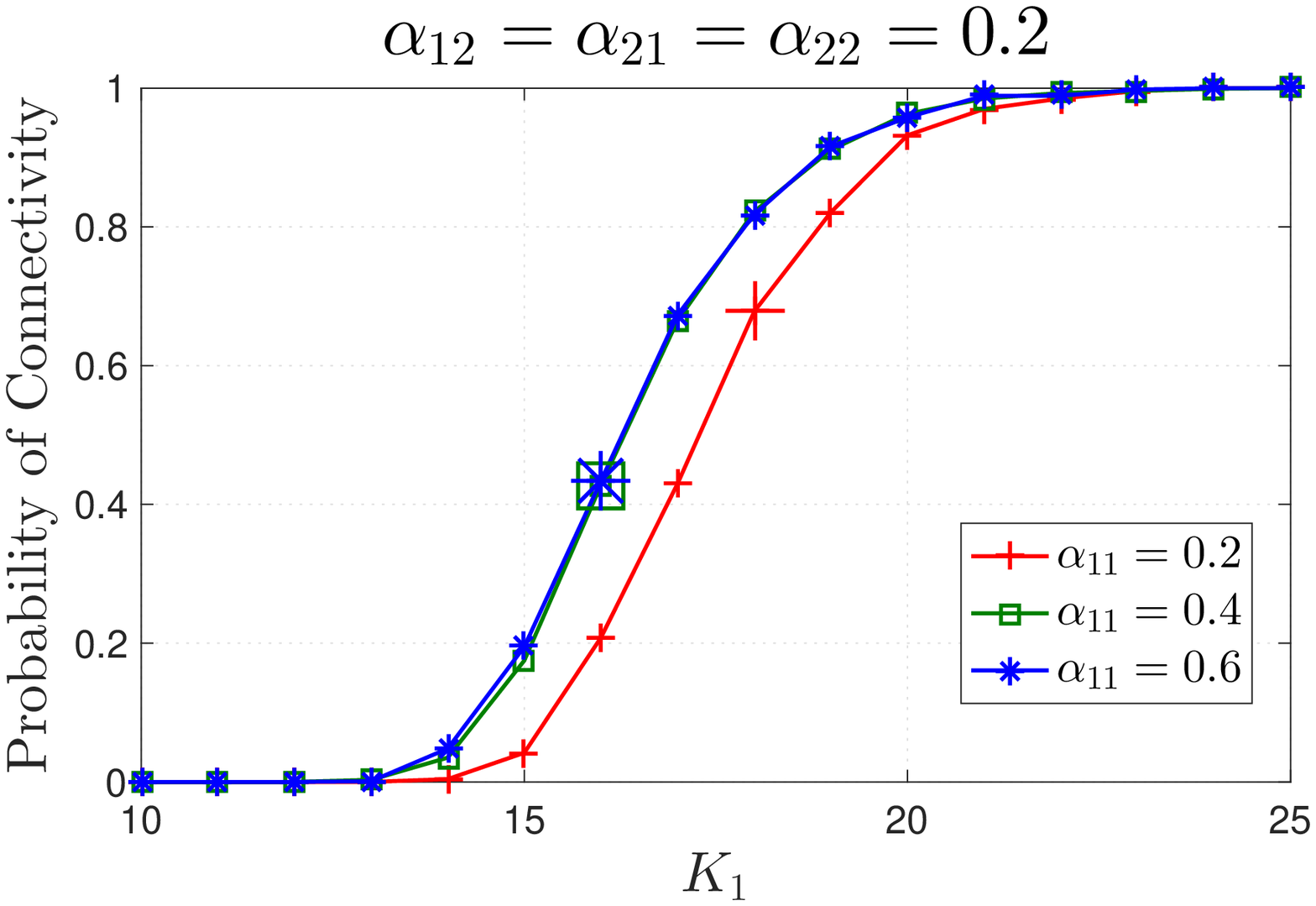}}
\caption{Empirical probability that $\mathbb{H}(n;\pmb{\mu},\pmb{\Theta})$ is connected as a function of $\pmb{K}$ for $\alpha_{11} = 0.2$, $\alpha_{11} = 0.4$, and $\alpha_{11} = 0.6$. We set $\alpha_{12}=\alpha_{22}=0.2$. Highlighted symbols stand for the critical threshold of connectivity asserted by Theorem~\ref{theorem:connectivity}.}
\label{fig:2}
\end{figure}

Next, we set the channel matrix to
\begin{equation*}
\pmb{\alpha}=
  \begin{bmatrix}
    \alpha_{11} & 0.2 \\
    0.2 & 0.2
  \end{bmatrix}
\end{equation*}
in Figure~\ref{fig:2}, and consider three different values for the parameter $\alpha_{11}$, namely, $\alpha_{11} = 0.2$, $\alpha_{11} = 0.4$, and $\alpha_{11} = 0.6$. We also vary $K_1$ from $10$ to $25$. The number of classes is fixed to $2$, with $\pmb{\mu}=\{0.5,0.5\}$. For each value of $K_1$, we set $K_2=K_1+5$. Similar to Figure~\ref{fig:1}, we obtain the empirical probability that $\mathbb{H}(n;\pmb{\mu},\pmb{\Theta})$ is connected versus $K_1$. As before, the critical threshold of connectivity asserted by Theorem~\ref{theorem:connectivity} is shown by highlighted symbols in each curve.

Note that for $\alpha_{11} \geq 0.4$, fixed $\alpha_{12}$, and fixed $\alpha_{22}$, the probability of connectivity (along with the critical value of $K_1$) behave in a similar fashion regardless of the particular value of $\alpha_{11}$. The reason behind this is intuitive. When $\alpha_{11}=0.2$, we have $\Lambda_m=\Lambda_1$, while for $\alpha_{11} \geq 0.4$, we have $\Lambda_m=\Lambda_2$. Consequently, the value of $\alpha_{11}$ (which only appears in $\Lambda_1$) becomes irrelevant to the scaling condition given by (\ref{eq:numerical_critical}).

\begin{figure}[t]
\centerline{\includegraphics[scale=0.6]{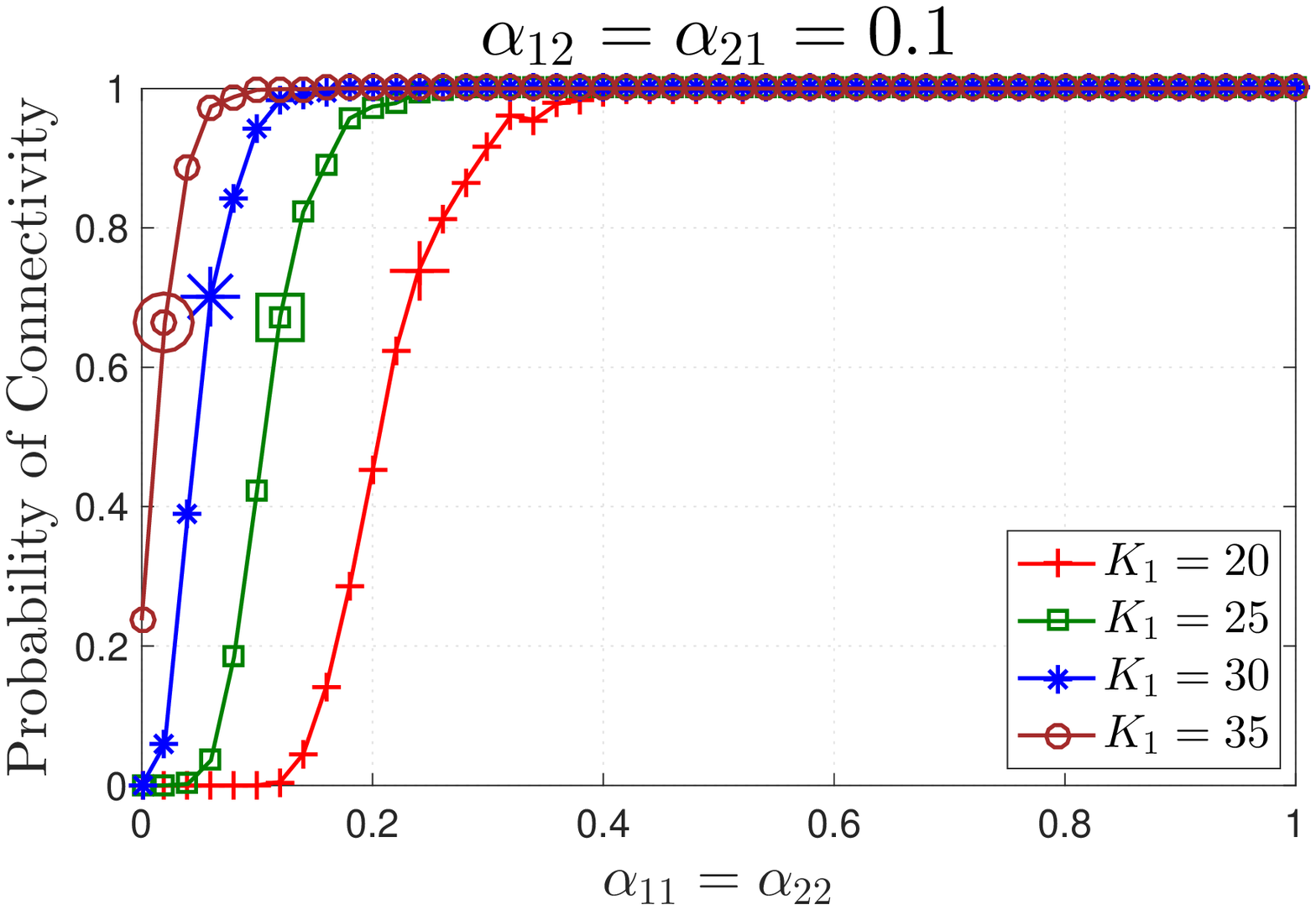}}
\caption{Empirical probability that $\mathbb{H}(n;\pmb{\mu},\pmb{\Theta})$ is connected as a function of $\alpha$ for $K_1 = 20$, $K_1 = 25$, $K_1 = 30$, and $K_1=35$. We set $\alpha_{12}=0.2$. Highlighted symbols stand for the critical threshold of connectivity asserted by Theorem~\ref{theorem:connectivity}.}
\label{fig:3}
\end{figure}

\begin{figure}[t]
\centerline{\includegraphics[scale=0.6]{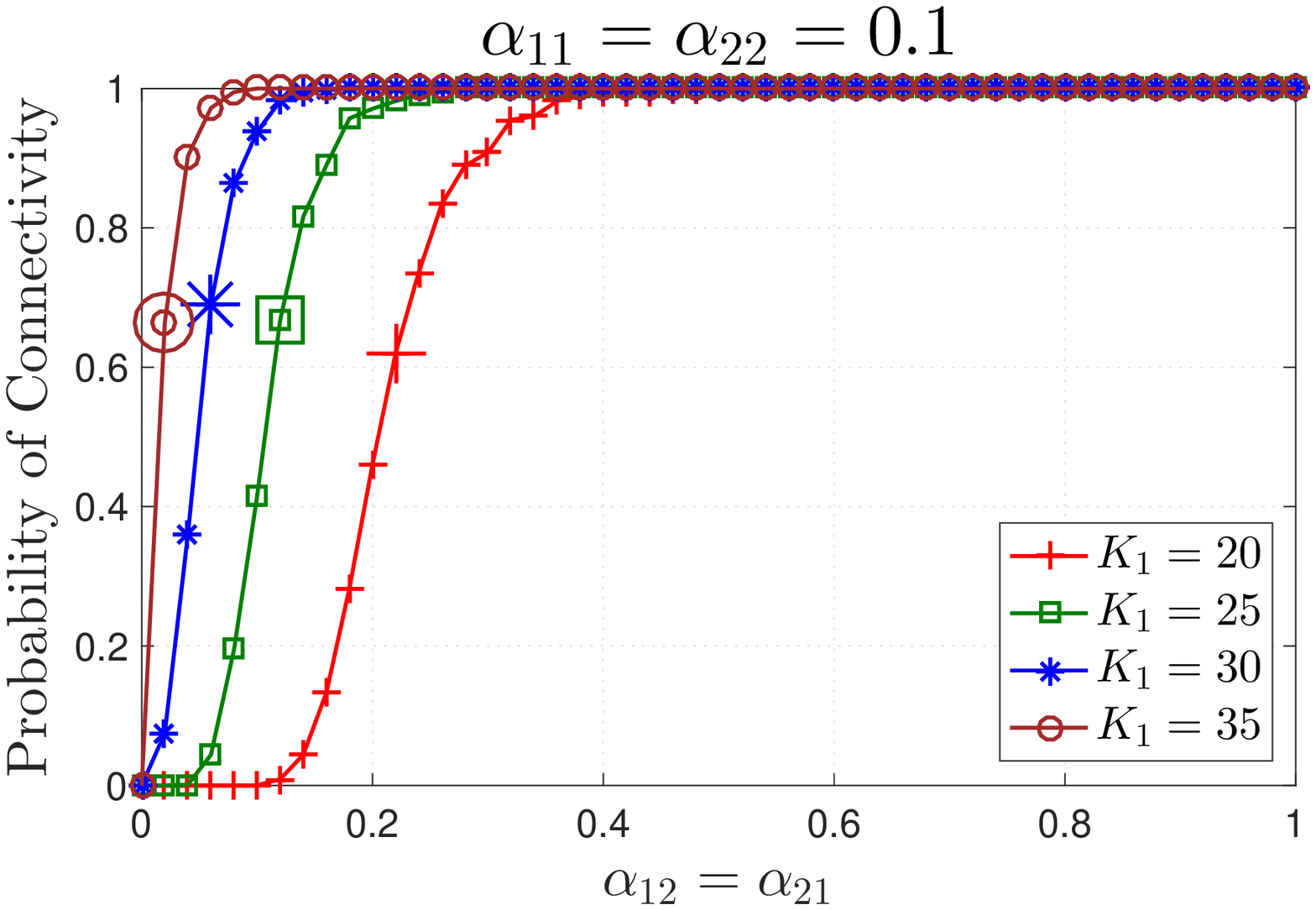}}
\caption{Empirical probability that $\mathbb{H}(n;\pmb{\mu},\pmb{\Theta})$ is connected as a function of $\alpha_{12}$ for $K_1 = 20$, $K_1 = 25$, $K_1 = 30$, and $K_1=35$. We set $\alpha_{11}=\alpha_{22}=0.2$. Highlighted symbols stand for the critical threshold of connectivity asserted by Theorem~\ref{theorem:connectivity}.}
\label{fig:4}
\end{figure}

Finally, we set the channel matrix to
\begin{equation*}
\pmb{\alpha}=
  \begin{bmatrix}
    \alpha & 0.2 \\
    0.2 & \alpha
  \end{bmatrix}
\end{equation*}
and consider four different values for the parameter $K_1$, namely, $K_1 = 20$, $K_1 = 25$, $K_1 = 30$, and $K_1=35$ while varying the parameter $\alpha$ from $0$ to $1$. The number of classes is fixed to $2$ with $\pmb{\mu}=\{0.5,0.5\}$ and we set $K_2=K_1+5$ for each value of $K_1$. We plot the empirical probability that $\mathbb{H}(n;\pmb{\mu},\pmb{\Theta})$ is connected versus $\alpha$ and highlight the critical threshold of connectivity asserted by Theorem~\ref{theorem:connectivity}. Note that $\mathbb{H}(n;\pmb{\mu},\pmb{\Theta})$ has a positive probability to be connected with $\alpha_{12}>0$ even when $\alpha=0$. In this case, the connected instances of $\mathbb{H}(n;\pmb{\mu},\pmb{\Theta})$ represent {\em connected bipartite graphs}, where one set of the bipartite graph represents class-$1$ nodes and the other represents class-$2$ nodes. The results given by Figure~\ref{fig:3} reveal the importance of cross-type edge probability in establishing a connected graph. In particular, when $\alpha_{11}=\alpha_{22}=0$, the graph could still be connected owing to cross-type edges. Indeed, the graph cannot be connected when cross-type edges have zero probability, even when same-type edges have positive probability since the graph would consist of at least two isolated components, as captured by Figure~\ref{fig:4}.

\section{Preliminaries}
Several technical results are collected here for convenience.
The first result follows easily from the scaling condition (\ref{scaling_condition_K}). 
\begin{prop} [{\cite[Proposition~4.1]{Yagan/Inhomogeneous}}]
For any scaling $K_1,K_2,\ldots,K_r,P:\mathbb{N}_0 \rightarrow \mathbb{N}_0^{r+1}$, we have
\begin{equation}
\lambda_1(n) \leq \lambda_2(n) \leq \ldots \leq \lambda_r(n)
\label{eq:isolated_ordering_of_lambda}
\end{equation}
for each $n=2,3,\ldots$.
\end{prop}

{\prop[{\cite[Proposition~4.4]{Yagan/Inhomogeneous}}]
For any set of positive integers $K_1,\ldots,K_r,P$ and any scalar $a \geq 1$, we have
\begin{equation}
\frac{\binom {P-\left \lceil{aK_i}\right \rceil }{K_j}}{\binom P{K_j}} \leq
\left(\frac{\binom {P-K_i}{K_j}}{\binom P{K_j}}\right)^a, \quad i,j=1,\ldots,r
\label{eq:isolated:combinatorial_bound}
\end{equation}
}

{\lemma
\label{lemma:isolated_1}
Consider a scaling $K_1,K_2,\ldots,K_r,P:\mathbb{N}_0 \rightarrow \mathbb{N}_0^{r+1}$ and a scaling $\pmb{\alpha}=\{\alpha_{ij}\}: \mathbb{N}_0 \rightarrow (0,1)^{r \times r}$ such that (\ref{eq:conn_K1r}) and (\ref{scaling_condition_KG_v2}) hold. We have
\begin{equation}
\alpha_{\min}(n) p_{1r}(n) = \Theta\left(\frac{\log n}{n}\right)
\label{eq:bound_on_alpha_min_p1r}
\end{equation}
}
\myproof
We note from (\ref{scaling_condition_KG_v2}) that
\begin{align}
\alpha_{mr}(n) p_{mr}(n) &\leq \frac{\Lambda_m(n)}{\mu_r} = \frac{c_n}{\mu_r} \frac{\log n}{n}, \nonumber
\end{align}

Next, we show that under (\ref{scaling_condition_K}), the quantity $p_{ij}(n)$ is increasing in both $i$ and $j$. Fix $n=2,3,\ldots$ and recall that under (\ref{scaling_condition_K}), $K_i$ increases as $i$ increases. For any $i,j$ such that $K_i + K_j > P$, we see from (\ref{eq:osy_edge_prob_type_ij}) that $p_{ij}(n)=1$; otherwise if $K_i + K_j \leq P$, we have $p_{ij}(n)<1$. Given that $K_i + K_j$ increases with both $i$ and $j$, it will be sufficient to show that $p_{ij}(n)$ increases with both $i$ and $j$ on the range where $K_i + K_j < P$. On that range, we have
\begin{equation}\nonumber
\frac{\binom{P-K_i}{K_j}}{\binom{P}{K_j}} = \prod_{\ell=0}^{K_i-1} \left(1-\frac{K_j}{P-\ell} \right)
\end{equation}
Hence, $\binom{P-K_i}{K_j} / \binom{P}{K_j}$ decreases with both $K_i$ and $K_j$, hence with $i$ and $j$. From (\ref{eq:osy_edge_prob_type_ij}), it follows that $p_{ij}(n)$ increases with $i$ and $j$. As a consequence, we have $p_{1r} \leq p_{mr}$ and it follows that
\begin{align}
\alpha_{\min}(n) p_{1r}(n) &\leq \alpha_{mr}(n) p_{mr}(n) \leq \frac{c_n}{\mu_r} \frac{\log n}{n}.
\label{eq:bound_on_alpha_min_p1r_part1}
\end{align}
Combining (\ref{eq:conn_K1r}) and (\ref{eq:bound_on_alpha_min_p1r_part1}) we readily obtain (\ref{eq:bound_on_alpha_min_p1r}).
\myendpf

{\lemma
\label{lemma:bounding_cl_event}
Consider a scaling $K_1,K_2,\ldots,K_r,P:\mathbb{N}_0 \rightarrow \mathbb{N}_0^{r+1}$ and a scaling $\pmb{\alpha}=\{\alpha_{ij}\}: \mathbb{N}_0 \rightarrow (0,1)^{r \times r}$ such that (\ref{scaling_condition_KG}) holds. From (\ref{eq:conn_K1r}), (\ref{eq:bounding_variance_of_K}), (\ref{eq:bounding_variance_of_alpha}), and (\ref{eq:bound_on_alpha_min_p1r_part1}), we have
\begin{equation}
\alpha_{\max}(n) p_{rr}(n) = o \left( \frac{\left(\log n \right)^{\tau+2}}{n} \right)
\label{eq:equation_bounding_cl_event}
\end{equation}
and
\begin{equation}
\alpha_{\min} p_{11}(n) =\omega\left(\frac{1}{n}\right),
\label{eq:conn_K1}
\end{equation}
}
\myproof
From (\ref{eq:bounding_variance_of_alpha}) and (\ref{eq:bound_on_alpha_min_p1r_part1}), we have
\begin{align}
\alpha_{\max}(n) p_{1r}(n) &= \left( \frac{\alpha_{\max}(n)}{\alpha_{\min}(n)}\right) \alpha_{\min}(n) p_{1r}(n) = O\left( \frac{\left(\log n \right)^{\tau+1}}{n} \right)
\end{align}
It is now immediate that  Lemma~\ref{lemma:bounding_cl_event} is established once we show that 
\begin{equation} 
\frac{p_{rr}(n)}{p_{1r}(n)} = o\left( \log n \right),
\label{eq:equation_bounding_cl_event_pt1}
\end{equation}
leading to
\begin{align} 
\alpha_{\max}(n) p_{rr}(n) &= \left( \frac{p_{rr}(n)}{p_{1r}(n)} \right) \alpha_{\max}(n) p_{1r}(n) = o\left( \frac{\left(\log n \right)^{\tau+2}}{n} \right) \nonumber
\end{align}
We proceed by establishing (\ref{eq:equation_bounding_cl_event_pt1}). The proof is similar with \cite[Lemma~5.4]{Rashad/Inhomo}, but we give it below for completeness. 

In particular, we will show that
\begin{align}
p_{rr}(n) &\leq \max \left( 2,  \frac{\log n}{w_n}\right) p_{1r}(n), \quad n=2,3, \ldots
\label{eq:to_show_p_rr_appendix}
\end{align}
 for some sequence $w_n$ such that $\lim_{n \to \infty} w_n = \infty$.
Fix $n=2,3,\ldots.$  We  have either
$p_{1r}(n) > \frac{1}{2}$,
or
$p_{1r}(n) \leq \frac{1}{2}$.
In the former case, it automatically holds that
\begin{equation}
p_{rr}(n) \leq 2 p_{1r}(n)
\label{eq:rig:1st_part}
\end{equation}
by virtue of the fact that $p_{rr}(n) \leq 1$.

Assume now that $p_{1r}(n) \leq \frac{1}{2}$.
We know from \cite[Lemmas~7.1-7.2]{yagan2012zero} that
\begin{equation}
1-e^{-\frac{K_{j,n}K_{r,n}}{P_n}} \leq p_{jr}(n) \leq \frac{K_{j,n}K_{r,n}}{P_n-K_{j,n}}, ~~ j=1,\ldots, r
\label{eq:rig_3}
\end{equation}
and it follows that
\begin{align}
\frac{K_{1,n}K_{r,n}}{P_n} \leq \log \left( \frac{1}{1-p_{1r}(n)} \right) \leq \log 2 < 1.
\label{eq:rig_4}
\end{align}
Using the fact that $1-e^{-x} \geq \frac{x}{2}$ with $x$ in $(0,1)$,
we then get
\begin{equation}
p_{1r}(n) \geq \frac{K_{1,n}K_{r,n}}{2P_n}.  
\label{eq:p_1r_lower_bound}
\end{equation}
In addition, using the upper bound in (\ref{eq:rig_3}) with $j=r$  gives
\[
p_{rr}(n) \leq  \frac{K_{r,n}^2}{P_n-K_{r,n}} \leq 2 \frac{K_{r,n}^2}{P_n}
\]
as we invoke (\ref{scaling_condition_K}). Combining the last two bounds we obtain
\begin{align}
\frac{p_{rr}(n)}{p_{1r}(n)} &\leq  4 \frac{K_{r,n}}{K_{1,n}}=4 \frac{\log n}{w_n}
\label{eq:bound_on_prr_by_p1r}
\end{align}
by virtue of (\ref{eq:bounding_variance_of_K}) for some sequence $w_n$ satisfying $\lim_{n \to \infty} w_n=\infty$.
Combining (\ref{eq:rig:1st_part}) and (\ref{eq:bound_on_prr_by_p1r}), we readily obtain (\ref{eq:to_show_p_rr_appendix}). This establishes (\ref{eq:equation_bounding_cl_event}).

Next, Combining (\ref{eq:conn_K1r}), and the fact that $p_{1r}(n) / p_{11}(n) = o(\log n)$ (see (\ref{eq:to_show_p_rr_appendix})), we get
\begin{align}
\alpha_{\min}(n) p_{11}(n) &= \left(\frac{p_{11}(n)}{p_{1r}(n)}\right) \alpha_{\min}(n) p_{1r}(n) = \omega \left( \frac{1}{n} \right) \nonumber
\end{align}
which readily establishes (\ref{eq:conn_K1}).
\myendpf

{
\lemma
\label{cor:new_corr}
Under (\ref{eq:conn_K1}), we have
\begin{equation}
\frac{K_{1,n}^2}{P_n}=\omega\left(\frac{1}{n \alpha_{\min}}\right),
\label{eq:conn_K1_new_trick}
\end{equation}
and
\begin{equation}
K_{1,n}=\omega(1).
\label{eq:conn_K1_new_trick_2}
\end{equation}
}
\myproof
It is a simple matter to check that 
$
p_{11}(n) \leq \frac{K_{1,n}^2}{P_n-K_{1,n}}
$; see \cite[Proposition~7.1-7.2]{yagan2012zero} for a proof. 
In view of (\ref{scaling_condition_K}) this gives $p_{11}(n) \leq 2 \frac{K_{1,n}^2}{P_n}$.
Thus, we have
 \begin{equation} \nonumber
\frac{K_{1,n}^2}{P_n}=\Omega\left( p_{11}(n) \right)=\omega\left(\frac{1}{n \alpha_{\min}}\right).
\end{equation}
From (\ref{eq:conn_Pn2}), (\ref{eq:conn_K1_new_trick}), and $\alpha_{\min} \leq 1$, we readily obtain (\ref{eq:conn_K1_new_trick_2}).
\myendpf

Other useful bound that will be used throughout is
\begin{align}
& (1 \pm x) \leq e^{\pm x}, \quad x \in (0,1)
\label{eq:isolated:exp_bound}
\\
& \binom{n}{\ell} \leq \left( \frac{en}{\ell} \right)^\ell, \quad \ell=1,\ldots,n, \quad n=1,2,\ldots
\label{eq:conn_bounds_1}
\\
& \sum_{\ell=2}^{\left \lfloor{\frac{n}{2}}\right \rfloor} \binom{n}{\ell} \leq 2^n
\label{eq:conn_bounds_2}
\end{align}

Finally, we find it useful to write
\begin{equation}
\log (1-x)=-x-\Psi(x), \quad x \in (0,1)
\label{eq:isolated_log_decomp}
\end{equation}
where  $\Psi(x)=\int_{0}^{x} \frac{t}{1-t} \ \text{dt}$.
From L'H\^{o}pital's Rule, we have
\begin{equation}
\lim_{x\to 0}  \frac{\Psi(x)}{x^2}=\frac{-x-\log (1-x)}{x^2}=\frac{1}{2}.
\label{eq:isolated_hopital}
\end{equation}

\section{Proof of Theorem~\ref{theorem:isolated_nodes}}
\label{sec:proof_isolated}
The proof of Theorem~\ref{theorem:isolated_nodes} relies on the method of first and second moments applied to the number of isolated nodes in $\mathbb{H}(n;\pmb{\mu},\pmb{\Theta}_n)$. Let $I_n(\pmb{\mu},\pmb{\Theta}_n)$ denote the total number of isolated nodes in $\mathbb{H}(n;\pmb{\mu},\pmb{\Theta}_n)$, namely,
\begin{equation}
I_n(\pmb{\mu},\pmb{\Theta}_n)=\sum_{\ell=1}^n \pmb{1}[v_\ell \text{ is isolated in }\mathbb{H}(n;\pmb{\mu},\pmb{\Theta}_n)]
\label{eq:isolated_In}
\end{equation}

The method of first moment \cite[Eqn. (3.1), p.
54]{JansonLuczakRucinski} gives
\begin{equation} \nonumber
1-\mathbb{E}[I_n(\pmb{\mu},\pmb{\Theta}_n)]\leq \mathbb{P}[I_n(\pmb{\mu},\pmb{\Theta}_n)=0] 
\end{equation}

\subsection{Establishing the one-law}
It is clear that in order to establish the one-law, namely that $ \lim_{n \to \infty} \mathbb{P}\left[  I_n(\pmb{\mu},\pmb{\Theta_n})=0\right]=1$, we need to show that
\begin{equation*}
\lim_{n \to \infty} \mathbb{E}[I_n(\pmb{\mu},\pmb{\Theta}_n)]=0.
\end{equation*}

Recalling (\ref{eq:isolated_In}), we have
\begin{align}
\mathbb{E}\left[I_n(\pmb{\mu},\pmb{\Theta}_n)\right]&=n \sum_{i=1}^r \mu_i \mathbb{P}\left[v_1 \text{ is isolated in }\mathbb{H}(n;\pmb{\mu},\pmb{\Theta}_n) \given[\big] t_1=i\right]\label{eq:isolated_independence_part1Rev} \\
&=n \sum_{i=1}^r \mu_i \mathbb{P}\left[\cap_{j=2}^n [v_j \nsim v_1] \mid t_1 = i \right]\nonumber \\
&=n \sum_{i=1}^r \mu_i \left(\mathbb{P}\left[v_2 \nsim v_1 \mid t_1 = i\right]\right)^{n-1} \label{eq:isolated_independence}
\end{align}
where (\ref{eq:isolated_independence_part1Rev}) follows by the exchangeability of the indicator functions appearing at (\ref{eq:isolated_In}) and (\ref{eq:isolated_independence}) follows by the conditional independence of the rvs $\{v_j \nsim v_1\}_{j=1}^n$ given $t_1$. By conditioning on the class of $v_2$, we find
\begin{align}
\mathbb{P}[v_2 \nsim v_1 \given[\big] t_1=i]
&=\sum_{j=1}^r \mu_j \mathbb{P}[v_2 \nsim v_1 \given[\big] t_1=i,t_2=j]=\sum_{j=1}^r \mu_j (1-\alpha p_{ij})=1-\Lambda_i(n).
\label{eq:isolated_OneLaw_first_step}
\end{align}
Using (\ref{eq:isolated_OneLaw_first_step}) in (\ref{eq:isolated_independence}), and recalling (\ref{eq:min_mean_degree}), (\ref{eq:isolated:exp_bound}) we obtain
\begin{align*}
\mathbb{E}[I_n(\pmb{\mu},\pmb{\Theta}_n)] &= n \sum_{i=1}^r \mu_i \left(1-\Lambda_i(n)\right)^{n-1}\nonumber \\
&\leq n \left(1-\Lambda_m(n)\right)^{n-1}\nonumber \\
&= n \left(1-c_n \frac{\log n}{n}\right)^{n-1}\nonumber \\
&\leq  e^{\log n \left(1-c_n  \frac{n-1}{n}\right)}
\end{align*}
Taking the limit as $n$ goes to infinity, we immediately get
\begin{equation*}
\lim_{n \to \infty} \mathbb{E}[I_n(\pmb{\mu},\pmb{\Theta}_n)]=0.
\end{equation*}
since $\lim_{n \to \infty} (1-c_n  \frac{n-1}{n})=1-c < 0$ under the enforced assumptions (with $c>1$) and the one-law is established.

\subsection{Establishing the zero-law}
Our approach in establishing the zero-law relies on the method of second moment applied to a variable that counts the number of nodes that are class-$m$ and isolated. Clearly if we can show that whp there exists at least one class-$m$ node that is isolated under the enforced assumptions (with $c<1$) then the zero-law would immediately follow.

Let $Y_n(\pmb{\mu},\pmb{\Theta}_n)$ denote the number of nodes that are class-$m$ and isolated in $\mathbb{H}(n;\pmb{\mu},\pmb{\Theta}_n)$, and let
\begin{equation} \nonumber
x_{n,i}(\pmb{\mu},\pmb{\Theta}_n)=\pmb{1}[t_i=m \cap v_i \text{ is isolated in }\mathbb{H}(n;\pmb{\mu},\pmb{\Theta}_n)],
\end{equation}
then we have $Y_n(\pmb{\mu},\pmb{\Theta}_n)=\sum_{i=1}^n x_{n,i}(\pmb{\mu},\pmb{\Theta}_n)$. By applying the method of second moments
\cite[Remark 3.1, p. 54]{JansonLuczakRucinski}  on $Y_n(\pmb{\mu},\pmb{\Theta}_n)$, we get
\begin{equation}
\mathbb{P}[Y_n(\pmb{\mu},\pmb{\Theta}_n)=0] \leq 1-\frac{ \left( \mathbb{E}[Y_n(\pmb{\mu},\pmb{\Theta}_n)] \right)^2}{\mathbb{E}[Y_n(\pmb{\mu},\pmb{\Theta}_n)^2]}  
\label{eq:isolated_ZeroLaw_bound}
\end{equation}
where
\begin{equation}
\mathbb{E}[Y_n(\pmb{\mu},\pmb{\Theta}_n)]=n \mathbb{E}[x_{n,1}(\pmb{\mu},\pmb{\Theta}_n)]
\label{eq:isolated_ZeroLaw_first_part}
\end{equation}
and
\begin{align}
\mathbb{E}[Y_n(\pmb{\mu},\pmb{\Theta}_n)^2]=&n \mathbb{E}[x_{n,1}(\pmb{\mu},\pmb{\Theta}_n)]+n(n-1)\mathbb{E}[x_{n,1}(\pmb{\mu},\pmb{\Theta}_n) x_{n,2}(\pmb{\mu},\pmb{\Theta}_n)]
\label{eq:isolated_ZeroLaw_second_part}
\end{align}
by exchangeability and the binary nature of the rvs $\{x_{n,i}(\pmb{\mu},\pmb{\Theta}_n) \}_{i=1}^n$.
Using (\ref{eq:isolated_ZeroLaw_first_part}) and  (\ref{eq:isolated_ZeroLaw_second_part}), we get
\begin{equation} \nonumber
\frac{\mathbb{E}[Y_n(\pmb{\mu},\pmb{\Theta}_n)^2]}{\left(\mathbb{E}[Y_n(\pmb{\mu},\pmb{\Theta}_n)]\right)^2}  = \frac{1}{n \mathbb{E}[x_{n,1}(\pmb{\mu},\pmb{\Theta}_n)]} + {\frac{n-1}{n} \frac{\mathbb{E}[x_{n,1}(\pmb{\mu},\pmb{\Theta}_n) x_{n,2}(\pmb{\mu},\pmb{\Theta}_n)]}{\left( \mathbb{E}[x_{n,1}(\pmb{\mu},\pmb{\Theta}_n)]\right)^2}}
\end{equation}

In order to establish the zero-law, we need to show that
\begin{equation*}
\lim_{n \to \infty} n \mathbb{E}[x_{n,1}(\pmb{\mu},\pmb{\Theta}_n)]= \infty,
\end{equation*}
and
\begin{equation}
\limsup_{n \to \infty} \left(\frac{\mathbb{E}[x_{n,1}(\pmb{\mu},\pmb{\Theta}_n) x_{n,2}(\pmb{\mu},\pmb{\Theta}_n)]}{\left(\mathbb{E}[x_{n,1}(\pmb{\mu},\pmb{\Theta}_n)]\right)^2}\right) \leq 1.
\label{eq:isolated_ZeroLaw_second_condition}
\end{equation}

{\prop
Consider a scaling $K_1,\ldots,K_r,P:\mathbb{N}_0 \rightarrow \mathbb{N}_0^{r+1}$ and a scaling $\pmb{\alpha}=\{\alpha_{ij}\}:=\mathbb{N}_0 \rightarrow (0,1)^{r \times r}$ such that (\ref{scaling_condition_KG}) holds with $\lim_{n \to \infty} c_n=c>0$. Then, we have
\begin{equation*}
\lim_{n \to \infty} n \mathbb{E}[x_{n,1}(\pmb{\mu},\pmb{\Theta}_n)]= \infty, \quad \text{if } c<1
\end{equation*}
}
\begin{proof}
We have
\begin{align}
n \mathbb{E}\left[x_{n,1}(\pmb{\mu},\pmb{\Theta}_n)\right] &=n \mathbb{E}\left[\pmb{1}[t_1=m \cap v_1 \text{ is isolated in }\mathbb{H}(n;\pmb{\mu},\pmb{\Theta}_n)]\right]\nonumber \\
&=n \mu_m \mathbb{P}\left[v_1 \text{ is isolated in }\mathbb{H}(n;\pmb{\mu},\pmb{\Theta}_n) \given[\big] t_1=m \right]\nonumber \\
&=n \mu_m \mathbb{P}\left[\cap_{j=2}^n [v_j \nsim v_1] \given[\big] t_1=m\right]\nonumber \\
&=n \mu_m \mathbb{P}\left[v_2 \nsim v_1\given[\big] t_1=m\right]^{n-1}\nonumber \\
&=n \mu_m \left(\sum_{j=1}^r \mu_j \mathbb{P}\left[v_2 \nsim v_1 \given[\big] t_1=1,t_2=j\right]\right)^{n-1}\nonumber \\
&=n \mu_m \left(\sum_{j=1}^r \mu_j (1-\alpha_{mj} p_{mj})\right)^{n-1} \label{eq:int_isol_prob_osy} \\
&=n \mu_m \left(1-\Lambda_m(n)\right)^{n-1} = \mu_m e^{\beta_n} 
\label{eq:isolated_ZeroLaw_simp1}
\end{align}
where 
\begin{equation*}
\beta_n=\log n+(n-1)\log (1-\Lambda_m(n)).\\
\end{equation*}
Recalling (\ref{eq:isolated_log_decomp}), we get
\begin{align}
\beta_n&=\log n-(n-1)\left(\Lambda_m(n)+\Psi(\Lambda_m(n))\right)\nonumber \\
&=\log n-(n-1)\left(c_n \frac{\log n}{n}+\Psi \left(c_n \frac{\log n}{n}\right)\right)\nonumber \\
&=\log n \left(1-c_n \frac{n-1}{n}\right) -(n-1) \left(c_n \frac{\log n}{n} \right)^2 \frac{\Psi \left(c_n \frac{\log n}{n}\right)}{\left(c_n \frac{\log n}{n}\right)^2}
 \label{eq:isolated_ZeroLaw_simp2}
\end{align}

Recalling (\ref{eq:isolated_hopital}), we have
\begin{equation}
\lim_{n \to \infty} \frac{\Psi \left(c_n \frac{\log n}{n}\right)}{\left(c_n \frac{\log n}{n}\right)^2} = \frac{1}{2}
\label{eq:isolated_ZeroLaw_simp3}
\end{equation}
since $c_n \frac{\log n}{n}=o(1)$. Thus, $\beta_n=\log n \left(1-c_n \frac{n-1}{n}\right)-o(1)$.
Using (\ref{eq:isolated_ZeroLaw_simp1}), (\ref{eq:isolated_ZeroLaw_simp2}), (\ref{eq:isolated_ZeroLaw_simp3}), and letting $n$ go to infinity, we get
\begin{equation*}
\lim_{n \to \infty} n \mathbb{E}[x_{n,1}(\pmb{\mu},\pmb{\Theta}_n)]= \infty
\end{equation*}
whenever $\lim_{n \to \infty} c_n=c < 1$.
\end{proof}

{\prop
Consider a scaling $K_1,\ldots,K_r,P:\mathbb{N}_0 \rightarrow \mathbb{N}_0^{r+1}$ and a scaling $\pmb{\alpha}=\{\alpha_{ij}\}:=\mathbb{N}_0 \rightarrow (0,1)^{r \times r}$ such that (\ref{scaling_condition_KG}) holds with $\lim_{n \to \infty} c_n=c>0$. Then, we have (\ref{eq:isolated_ZeroLaw_second_condition}) if $c<1$.
\label{prop:new_osy}
}
\begin{proof}
Consider fixed $\pmb{\Theta}$. 
\begin{align*}
\mathbb{E}\left[x_{n,1}(\pmb{\mu},\pmb{\Theta}) x_{n,2}(\pmb{\mu},\pmb{\Theta})\right] &=\mathbb{E}\left[\pmb{1}[v_1 \text{ is isolated }, v_2 \text{ is isolated} \cap t_1=m,t_2=m]\right]\nonumber \\
&=\mu_m^2 \mathbb{E}\left[\pmb{1}[v_1 \text{ is isolated }, v_2 \text{ is isolated}] \given[\Big]  t_1=m,t_2=m\right]\nonumber \\
&=\mu_m^2 \mathbb{E} \left[\pmb{1}[v_1 \nsim v_2]  \prod_{k=3}^n \pmb{1}[v_k \nsim v_1, v_k \nsim v_2] \given[\right] t_1=t_2=m\Bigg]
\end{align*}
Now we condition on $\Sigma_1 $ and $\Sigma_2$ and 
note that i) $\Sigma_1$ and $\Sigma_2$  determine $t_1$ and $t_2$;
and ii) the events $[v_1 \nsim v_2], \{[v_k \nsim v_1 \cap v_k \nsim v_2]\}_{k=3}^n$ are mutually independent given
$\Sigma_1$ and $\Sigma_2$.
Thus, we have
\begin{align} 
\mathbb{E}\left[x_{n,1}(\pmb{\mu},\pmb{\Theta}) x_{n,2}(\pmb{\mu},\pmb{\Theta})\right]= \mu_m^2 \mathbb{E}\left[\mathbb{P}\left[v_1 \nsim v_2 \given[\Big] \Sigma_1,\Sigma_2\right]  \times \prod_{k=3}^n \mathbb{P}\left[v_k \nsim v_1 \cap v_k \nsim v_2 \given[\Big] \Sigma_1,\Sigma_2\right]\given[\bigg] t_1=t_2=m \right]
 \label{eq:osy_new_zero_isol}
 \end{align}
 
Define the $\{0,1\}$-valued rv $u(\pmb{\theta})$ by
\begin{equation}
u(\pmb{\theta}):=\pmb{1}[\Sigma_1 \cap \Sigma_2 \neq \emptyset].
\label{eq:isolated_u(theta)}
\end{equation}
Next, with $\ell=1,2,\ldots,n-1$, define $\nu_{\ell,j}(\pmb{\alpha})$ by
\begin{align}
 \nu_{\ell,j}(\pmb{\alpha}) :=\{i=1,2,\ldots,\ell : B_{ij}(\pmb{\alpha})=1 \} 
\label{eq:isolated_v(alpha)}
\end{align}
for each $j=\ell+1,\ldots,n$. Namely, $\nu_{\ell,j}(\pmb{\alpha})$ is the set of nodes in $\{1,\ldots,\ell\}$ that are adjacent to node $j$ in $\mathbb{G}(n;\pmb{\mu},\pmb{\alpha})$. With these definitions in mind, (\ref{eq:osy_new_zero_isol}) gives
\begin{align*}
\mathbb{E}[x_{n,1}(\pmb{\mu},\pmb{\Theta}) x_{n,2}(\pmb{\mu},\pmb{\Theta})] = \mu_m^2 \mathbb{E}\vast[(1-\alpha_{mm})^{u(\pmb{\theta})}\mathlarger{\mathlarger{\prod}}_{k=3}^n \frac{\dbinom {P-\left|\cup_{i \in \nu_{2,k}(\pmb{\alpha})} \Sigma_i\right|}{|\Sigma_k|}}{\dbinom P{|\Sigma_k|}} \given[\vast] t_1=t_2=m \vast] \nonumber 
\end{align*}

Conditioned on $u(\pmb{\theta})=0$ and $v_1, v_2$ being class-$m$, we have
\begin{equation*} 
\left|\cup_{i \in \nu_{2,m}(\pmb{\alpha})} \Sigma_i \right|=\left|\nu_{2,k}(\pmb{\alpha})\right|K_m.
\end{equation*}
Also, we have
\[
\mathbb{P}[u(\pmb{\theta_n}) = 0 \given[] t_1=t_2=m] = 1-p_{mm}.
\]
Thus, we get
\begin{align}
&\mathbb{E}\left[x_{n,1}(\pmb{\mu},\pmb{\Theta}) x_{n,2}(\pmb{\mu},\pmb{\Theta}) ~ \pmb{1}[u(\pmb{\theta})=0]\right] \nonumber \\
&=\mu_m^2 (1-p_{mm}) \mathbb{E}\left[\mathlarger{\mathlarger{\prod}}_{k=3}^n \frac{\dbinom {P-| \nu_{2,k}(\pmb{\alpha})  K_m |}{|\Sigma_k|}}{\dbinom P{|\Sigma_k|}} \given[\Bigg]t_1=t_2=m \right]\nonumber \\
&=\mu_m^2 (1-p_{mm}) \mathbb{E}\left[\frac{\dbinom {P-|\nu_{2,3}(\pmb{\alpha})| K_m}{|\Sigma_3|}}{\dbinom P{|\Sigma_3|}}\given[\Bigg]t_1=t_2=m \right]^{n-2}
\label{eq:RevIndep} \\
& = \mu_m^2 (1-p_{mm})
 \left(\mathlarger{\mathlarger{\sum}}_{j=1}^r \mu_j \mathbb{E}\left[\frac{\dbinom {P-|\nu_{2,3}(\pmb{\alpha})| K_m}{|\Sigma_3|}}{\dbinom P{|\Sigma_3|}} \given[\vast] 
\begin{array}{c}
 t_1=t_2=m \\
 t_3 =j
\end{array}
 \right]\right)^{n-2}
 \nonumber \\
& \leq \mu_m^2 (1-p_{mm})  \left(\mathlarger{\mathlarger{\sum}}_{j=1}^r \mu_j  \mathbb{E}\left[ \left(\frac{\dbinom {P-K_m}{K_j}}{\dbinom P{K_j}}\right)^{|\nu_{2,3}(\pmb{\alpha})|} \given[\vast] 
\begin{array}{c}
 t_1=t_2=m \\
 t_3 =j
\end{array} \right] \right)^{n-2},
\end{align}
where we use (\ref{eq:isolated:combinatorial_bound}) in the last step. Note that conditioned on $t_1=t_2=m$, the random variables $\{\left| \nu_{2,k}(\pmb{\alpha}) \right|\}_{k=3}^n$ are independent and identically distributed, hence (\ref{eq:RevIndep}) follows. In particular
\begin{equation}\nonumber
\left| \nu_{2,k}(\pmb{\alpha}) \right| \given[\Big] t_1=t_2=m \sim \text{Binomial}\left(2,\alpha_{mj} \right) \quad \text{with probability } \mu_j, \quad k=3,4,\ldots,n
\end{equation}
The above distributional equality could be explained as follows. We may write $\left| \nu_{2,k}(\pmb{\alpha}) \right| = \pmb{1} \left[v_1 \sim_G v_k \right] +  \pmb{1} \left[v_2 \sim_G v_k \right]$. Observe that conditioned on $t_1=t_2=m$, we know that nodes $v_1$ and $v_2$ belong to class-$m$ in $\mathbb{G}\left( n; \pmb{\mu}, \pmb{\alpha}\right)$. If node $v_k$ is class-$j$ (an event that has probability $\mu_j$), then $\pmb{1} \left[v_1 \sim_G v_k \right]$ and $\pmb{1} \left[v_2 \sim_G v_k \right]$ are each distributed as Bernoulli random variable with parameter $\alpha_{mj}$.

Now, let
\begin{equation}
Z_j=\frac{\dbinom {P-K_m}{K_j}}{\dbinom P{K_j}}=1-p_{mj}, \quad j=1,\ldots,r.
\label{eq:Z_teta_defn}
\end{equation}
Then,
\begin{align}
\mathbb{E}[x_{n,1}(\pmb{\mu},\pmb{\Theta}) x_{n,2}(\pmb{\mu},\pmb{\Theta}) \pmb{1}\left[u(\pmb{\theta})=0\right]] \leq \mu_m^2 (1-p_{mm}) \left(\sum_{j=1}^r \mu_j \mathbb{E}\left[Z_j^{|\nu_{2,3}(\pmb{\alpha})|} \given[\bigg] 
\begin{array}{c}
 t_1=t_2=m \\
 t_3 =j
\end{array}\right] \right)^{n-2}
\label{eq:isolated_second_moment_6}
\end{align}

Note that 
\begin{equation*}
\left|\nu_{2,3}(\pmb{\alpha})\right| \given[\bigg] \begin{array}{c}
t_1=t_2=m\\
t_3=j 
\end{array}
\sim \mathrm{Binomial}(2,\alpha_{mj})
\end{equation*}
Hence,
\begin{align}
\mathbb{E}\left[Z_j^{|\nu_{2,3}(\pmb{\alpha})|}\given[\bigg] 
\begin{array}{c}
 t_1=t_2=m \\
 t_3 =j
\end{array} \right] &=\sum_{i=0}^2 \dbinom{2}{i} \alpha_{mj}^i (1-\alpha_{mj})^{2-i} Z_j^i \nonumber \\
&=\sum_{i=0}^2 \dbinom{2}{i} \alpha_{mj}^i (1-\alpha_{mj})^{2-i} \left(1-p_{mj} \right)^i \nonumber \\
&=1-2\alpha_{mj} p_{mj}+\left( \alpha_{mj}p_{mj} \right)^2
\end{align}
upon recalling (\ref{eq:Z_teta_defn}). Next, let $W$ be a rv that takes the value $\alpha_{mj} p_{mj}$ with probability $\mu_j$. It follows that
\begin{align}
\sum_{j=1}^r \mu_j \mathbb{E}\left[Z_j^{|\nu_{2,3}(\pmb{\alpha})|} \given[\bigg] 
\begin{array}{c}
 t_1=t_2=m \\
 t_3 =j
\end{array}\right]=1-2\Lambda_m+\sum_{j=1}^r \mu_j \left( \alpha_{mj}p_{mj} \right)^2 =1-2\Lambda_m+\mathbb{E}\left[ W^2 \right] \nonumber
\end{align}
Next, we recall (\ref{eq:HetER_s}) and let
\begin{align*}
&k:=\arg \min_j \alpha_{mj} p_{mj} 
\end{align*}
Now, in view of Popoviciu's inequality \cite[pp.~9]{jensen_99}, we see that 
\begin{align}
\mathrm{var} (W) &\leq \frac{1}{4} \left( W_{\max}-W_{\min} \right)^2 \nonumber \\
&=\frac{1}{4} \left( \alpha_{ms} p_{ms}-\alpha_{mk} p_{mk} \right)^2 \nonumber \\
&\leq \frac{1}{4} \left( \alpha_{ms} p_{ms} \right)^2 
\label{eq:HetER1}
\end{align}
We also know from (\ref{eq:osy_mean_edge_prob_in_system}) that
\begin{equation}
\alpha_{ms} p_{ms} \leq \frac{1}{\mu_s} \Lambda_m
\label{eq:HetER2}
\end{equation}
From (\ref{eq:HetER1}) and (\ref{eq:HetER2}), we get
\begin{equation}
\mathrm{var} (W) \leq \frac{1}{4\mu_s^2} \Lambda_m^2
\end{equation}
It is now immediate that
\begin{align}
\mathbb{E}\left[ W^2 \right] &= \left( \mathbb{E}\left[ W\right] \right)^2 + \mathrm{var} (W) \leq \left( 1 + \frac{1}{4 \mu_s^2}\right) \Lambda_m^2 
\label{eq:HetER3}
\end{align}
by virtue of the fact that $\mathbb{E}\left[ W\right] =\Lambda_m$. Using (\ref{eq:HetER3}) into (\ref{eq:isolated_second_moment_6}), we readily obtain
\begin{align}
\mathbb{E}[x_{n,1}(\pmb{\mu},\pmb{\Theta}) x_{n,2}(\pmb{\mu},\pmb{\Theta}) \pmb{1}\left[u(\pmb{\theta})=0\right]]  \leq \mu_m^2 (1-p_{mm}) \left(1-2\Lambda_m +\left( 1 + \frac{1}{4 \mu_s^2}\right) \Lambda_m^2  \right)^{n-2}
\label{eq:HetER4}
\end{align}

Next, conditioning on $u(\pmb{\theta})=1$ and $t_1=t_2=m$, we have
\begin{align*}
|\cup_{i \in \nu_{2,k}(\pmb{\alpha})} \Sigma_i|  & =
\begin{cases} 
       0    \hfill & \text{ if $|\nu_{2,k}(\pmb{\alpha})|=0$} \\
       K_m \hfill & \text{ if $|\nu_{2,k}(\pmb{\alpha})|=1$} \\
       2K_m-|\Sigma_1 \cap \Sigma_2| \hfill & \text{ if $|\nu_{2,k}(\pmb{\alpha})|=2$} \\
\end{cases} \nonumber
\end{align*}
and by a crude bounding argument, we have
\begin{equation}
|\cup_{i \in \nu_{2,k}(\pmb{\alpha})} \Sigma_i| \geq K_m \pmb{1}[|\nu_{2,k}(\pmb{\alpha})|>0]
\label{eq:isolated_second_moment_5}
\end{equation}
Using (\ref{eq:isolated_second_moment_5}) and recalling the analysis for $\mathbb{E}[x_{n,1}(\pmb{\mu},\pmb{\Theta}) x_{n,2}(\pmb{\mu},\pmb{\Theta}) \pmb{1}[u(\pmb{\theta})=0]]$, we obtain
\begin{align}
\mathbb{E}[x_{n,1}(\pmb{\mu},\pmb{\Theta}) x_{n,2}(\pmb{\mu},\pmb{\Theta}) \pmb{1}[u(\pmb{\theta})=1]]\leq \mu_m^2 (1-\alpha_{mm})p_{mm}  \Bigg( \sum_{j=1}^r \mu_j \mathbb{E}\Bigg[Z_j^{\pmb{1}[|\nu_{2,3}(\pmb{\alpha})|>0]} \given[\Bigg] 
\begin{array}{c}
 t_1=t_2=m \\
 t_3 =j
\end{array} \Bigg] \Bigg)^{n-2} 
\label{eq:isolated_second_moment_8}
\end{align} 
where
\begin{align*}
\mathbb{E}\left[Z_j^{\pmb{1}[|\nu_{2,3}(\pmb{\alpha})|>0]} \given[\bigg] 
\begin{array}{c}
 t_1=t_2=m \\
 t_3 =j
\end{array}\right]=\left(1-\alpha_{mj} \right)^2 + \left( 1- \left(1-\alpha_{mj} \right)^2 \right) Z_j =1-2\alpha_{mj} p_{mj} + \alpha_{mj}^2 p_{mj}
\end{align*}
and it follows that
\begin{align}
 \sum_{j=1}^r \mu_j \mathbb{E}\left[Z_j^{\pmb{1}[|\nu_{2,3}(\pmb{\alpha})|>0]} \given[\bigg] 
\begin{array}{c}
 t_1=t_2=m \\
 t_3 =j
\end{array}\right]&=1-2\Lambda_m + \sum_{j=1}^r \mu_j \alpha_{mj}^2 p_{mj} \nonumber \\
 &\leq 1-2\Lambda_m + \alpha_{md} \sum_{j=1}^r \mu_j \alpha_{mj} p_{mj} \nonumber \\
 &= 1- \left(2-\alpha_{md} \right) \Lambda_m 
\label{eq:HetER5}
\end{align}
upon recalling (\ref{eq:HetER0}). From (\ref{eq:isolated_second_moment_8}) and (\ref{eq:HetER5}), we readily obtain
\begin{align}
\mathbb{E}[x_{n,1}(\pmb{\mu},\pmb{\Theta}) x_{n,2}(\pmb{\mu},\pmb{\Theta}) \pmb{1}\left[u(\pmb{\theta})=1\right]] \leq \mu_m^2 (1-\alpha_{mm}) p_{mm} \left(1- \left(2-\alpha_{md} \right) \Lambda_m \right)^{n-2}
\label{eq:HetER6}
\end{align}

Combining (\ref{eq:HetER4}) and (\ref{eq:HetER6}), we get
\begin{align}
\mathbb{E}[x_{n,1}(\pmb{\mu},\pmb{\Theta}) x_{n,2}(\pmb{\mu},\pmb{\Theta})] &= \mathbb{E}[x_{n,1}(\pmb{\mu},\pmb{\Theta}) x_{n,2}(\pmb{\mu},\pmb{\Theta}) \left(\pmb{1}[u(\pmb{\theta})=0] + \pmb{1}[u(\pmb{\theta})=1]  \right)]\nonumber \\
& \leq \mu_m^2 (1-p_{mm}) \left(1-2\Lambda_m +\left( 1 + \frac{1}{4 \mu_s^2}\right) \Lambda_m^2  \right)^{n-2} \nonumber \\
& \quad + \mu_m^2 (1-\alpha_{mm}) p_{mm} \left(1- \left(2-\alpha_{md} \right) \Lambda_m \right)^{n-2}
\label{eq:isolated_second_moment_9}
\end{align}

It is also clear that
\begin{align}
\mathbb{E}[x_{n,1}(\pmb{\mu},\pmb{\Theta})]&=\mu_m \left(1-\Lambda_m \right)^{n-1}
\label{eq:isolated_second_moment_10}
\end{align}
Combining (\ref{eq:isolated_second_moment_9}) and (\ref{eq:isolated_second_moment_10}), we get
\begin{align}
\frac{\mathbb{E}[x_{n,1}(\pmb{\mu},\pmb{\Theta}) x_{n,2}(\pmb{\mu},\pmb{\Theta})] }{\mathbb{E}[x_{n,1}(\pmb{\theta})]^2} 
& \leq  (1-p_{mm})  \frac{ \left( 1-2\Lambda_m + \left( 1+ \frac{1}{4 \mu_s^2}\right) \Lambda_m^2\right)^{n-2}}{\left(1- \Lambda_m \right)^{2(n-1)}} +  p_{mm} \frac{ \left( 1-2\Lambda_m+\alpha_{md}\Lambda_m\right)^{n-2}}{ \left( 1-\Lambda_m\right)^{2(n-1)}} \nonumber \\
&:=A+B
\label{eq:defn_AB}
\end{align}
where we use the fact that $1-\alpha_{mm} \leq 1$.

We now consider a scaling $\pmb{\Theta} : \mathbb{N}_0 \rightarrow \mathbb{N}_0^{r+1} \times (0,1)^{r \times r}$
as stated in Proposition \ref{prop:new_osy} and
bound the terms $A$ and $B$ in turn. Our goal is to show that
\begin{equation}
\limsup_{n \to \infty} (A+B) \leq 1.
\label{eq:to_show_last_step_zero_law}
\end{equation}

We have
\begin{align*}
A &= \frac{1-p_{mm}}{\left( 1- \Lambda_m\right)^2} \left( 1+ \frac{1}{4 \mu_s^2} \left( \frac{\Lambda_m}{1-\Lambda_m} \right)^2 \right)^{n-2}  \leq \frac{1-p_{mm}}{\left( 1- \Lambda_m\right)^2} e^{\rho_n}
\end{align*}
where
\begin{align*}
\rho_n &\leq \left( \frac{c_n}{2 \mu_s} \right)^2 n \left( \frac{\log n}{n-c_n \log n}\right)^2 =o(1)\nonumber 
\end{align*}
and
\begin{equation}
 \left(1-\Lambda_m(n) \right)^2=1-o(1) 
\label{eq:osy_second_moment_new_EZ}
\end{equation}
since 
$\Lambda_m(n) = c_n \log n / n$. 
Thus, we have
\begin{equation}
A \leq \left(1-p_{mm} \right) \left( \left(1+o(1) \right) e^{o(1)} \right)
\label{eq:bound_on_A}
\end{equation}

We now consider the second term  in (\ref{eq:defn_AB}). Recall (\ref{eq:osy_second_moment_new_EZ}), we have
\begin{align}
B &= \frac{p_{mm}}{\left( 1- \Lambda_m\right)^2} \left( 1+\frac{\Lambda_m \left( \alpha_{md} - \Lambda_m \right)}{\left( 1- \Lambda_m\right)^2}\right)^{n-2}\leq \frac{p_{mm}}{\left( 1- \Lambda_m\right)^2}  e^{\psi_n} \nonumber
\end{align}
Now, recalling (\ref{scaling_condition_KG_v2}), we get
\begin{align}
\psi_n &\leq n \frac{\Lambda_m \left( \alpha_{md} - \Lambda_m \right)}{\left( 1- \Lambda_m\right)^2}= \frac{c_n \alpha_{md} \log n}{ \left( 1- c_n \frac{\log n}{n}\right)^2} - \frac{c_n^2 \frac{\left(\log n \right)^2}{n}}{ \left( 1- c_n \frac{\log n}{n}\right)^2} = \frac{c_n \alpha_{md} \log n}{ \left( 1- c_n \frac{\log n}{n}\right)^2} - o(1) \nonumber
\end{align}
Thus, we have
\begin{equation}
B \leq p_{mm} . \exp \left( \frac{c_n \alpha_{md} \log n}{ \left( 1- c_n \frac{\log n}{n}\right)^2} \right) . \left( \left(1+o(1) \right) e^{o(1)} \right)
\label{eq:bound_on_B}
\end{equation}

We will now establish the desired result (\ref{eq:to_show_last_step_zero_law})
by using (\ref{eq:bound_on_A}) and (\ref{eq:bound_on_B}). Our approach is to consider the cases i) $\lim_{n \to \infty} \alpha_{md}(n) \log n = 0 $ and ii) $\lim_{n \to \infty} \alpha_{mm}(n) \log n \in (0, \infty]$ separately.

\paragraph{Assume that $\lim_{n \to \infty} \alpha_{md}(n) \log n=0$}. From  (\ref{eq:bound_on_B}) we get
$B \leq (1+o(1))  p_{mm}$ and upon using (\ref{eq:bound_on_A}) we see that 
$A+B \leq  (1+o(1))$ establishing (\ref{eq:to_show_last_step_zero_law}) along subsequences with $\lim_{n \to \infty} \alpha_{md}(n) \log n=0$.

\paragraph{Assume that $\lim_{n \to \infty} \alpha_{mm}(n) \log n \in (0,\infty]$}. From (\ref{eq:osy_mean_edge_prob_in_system}), we have
\begin{align*} 
\Lambda_m=\sum_{j=1}^r \mu_j \alpha_{mj} p_{mj} \geq \mu_m \alpha_{mm} p_{mm}
\end{align*}
Thus,
\begin{align}
B &\leq \frac{1}{\mu_m} \frac{\Lambda_m}{\alpha_{mm}} . \exp \left( \frac{c_n \alpha_{md} \log n}{ \left( 1- c_n \frac{\log n}{n}\right)^2} \right)  = \frac{1}{\mu_m} \Lambda_m \log n . \frac{\exp \left( \frac{c_n \alpha_{md} \log n}{ \left( 1- c_n \frac{\log n}{n}\right)^2} \right)}{\alpha_{mm} \log n} \leq \frac{1}{\mu_m} c_n \left( \log n\right)^2 \frac{n^{-1+\frac{c_n}{\left( 1- c_n \frac{\log n}{n}\right)^2}}}{\alpha_{mm} \log n} \nonumber
\end{align}
since $\alpha_{md} \leq 1$. We note that
\begin{equation*}
\lim_{n \to \infty} {-1+\frac{c_n}{\left( 1- c_n \frac{\log n}{n}\right)^2}}=-1+c<0
\end{equation*}
for $c<1$. Thus, it follows that $B = o(1)$ upon noting that $\lim_{n \to \infty} \alpha_{mm} \log n = \alpha_{*} \in (0,\infty]$. From (\ref{eq:bound_on_A}) and the fact that $p_{mm} \leq 1$, we have $A+B \leq 1+o(1)$, and
(\ref{eq:to_show_last_step_zero_law}) follows. 

Note that if the matrix $\pmb{\alpha}$ is designed in such a way that $\alpha_{ii} = \max_j \{\alpha_{ij}\}$, i.e., two nodes of the same type are more likely to be adjacent in $\mathbb{G}(n;\pmb{\mu}, \pmb{\alpha})$, then we have $\alpha_{md} = \alpha_{mm}$ and the above two cases collapse to i) $\lim_{n \to \infty} \alpha_{mm}(n) \log n = 0 $ or ii) $\lim_{n \to \infty} \alpha_{mm}(n) \log n \in (0, \infty]$. At this point, the zero-law follows even when the sequence $\alpha_{mm} \log n$ does not have a limit by virtue of the {\em subsubsequence principle} \cite[p. 12]{JansonLuczakRucinski} (see also \cite[Section~7.3]{EletrebyKConn}). In other words, if $\alpha_{md} = \alpha_{mm}$, then the zero-law follows without any conditions on the sequence $\alpha_{mm}(n) \log n $. 
\end{proof}

\section{Proof of Theorem~\ref{theorem:connectivity}}
Let $C_n(\pmb{\mu},\pmb{\Theta}_n)$ denote the event that the graph $\mathbb{H}(n,\pmb{\mu},\pmb{\Theta}_n)$ is connected,
and with a slight abuse of notation, let $I_n(\pmb{\mu},\pmb{\Theta}_n)$ denote the event that the graph $\mathbb{H}(n,\pmb{\mu},\pmb{\Theta}_n)$ has no isolated nodes.
It is clear that if a random graph is connected then it does not have any isolated node, hence
\begin{equation*}
C_n(\pmb{\mu},\pmb{\Theta}_n) \subseteq I_n(\pmb{\mu},\pmb{\Theta}_n)
\end{equation*}
and we get
\begin{equation}
\mathbb{P}[C_n(\pmb{\mu},\pmb{\Theta}_n)] \leq \mathbb{P}[I_n(\pmb{\mu},\pmb{\Theta}_n)]
\label{eq:conn_ZeroLaw}
\end{equation}
and
\begin{align} \label{eq:conn_OneLaw}
\mathbb{P}[C_n(\pmb{\mu},\pmb{\Theta}_n)^c] & = \mathbb{P}[I_n(\pmb{\mu},\pmb{\Theta}_n)^c]+\mathbb{P}[C_n(\pmb{\mu},\pmb{\Theta}_n)^c \cap I_n(\pmb{\mu},\pmb{\Theta}_n)].
\end{align}

In view of (\ref{eq:conn_ZeroLaw}), we obtain the zero-law for connectivity, i.e., that
\begin{equation} \nonumber
\lim_{n\to\infty} \mathbb{P}[\mathbb{H}(n;\pmb{\mu},\pmb{\Theta}_n) \text{ is connected}]= 0    \quad \text{ if } \quad c<1,
\end{equation}
immediately from the zero-law part of
Theorem \ref{theorem:isolated_nodes}, i.e., from that 
$\lim_{n\to\infty} \mathbb{P}[I_n(\pmb{\mu},\pmb{\Theta}_n)]=0$ if $c<1$ under the enforced assumptions.
It remains to establish the one-law for connectivity. In the remainder of this section, we assume that (\ref{scaling_condition_KG}) holds for some $c>1$. 
From Theorem \ref{theorem:isolated_nodes} and (\ref{eq:conn_OneLaw}), we see that the one-law for connectivity, i.e., that
\[
\lim_{n\to\infty} \mathbb{P}[\mathbb{H}(n;\pmb{\mu},\pmb{\Theta}_n) \text{ is connected}]= 1    \quad \text{ if } \quad c > 1,
\]
will follow if we show that
\begin{align}
\lim_{n \to \infty} \mathbb{P}[C_n(\pmb{\mu},\pmb{\Theta}_n)^c \cap I_n(\pmb{\mu},\pmb{\Theta}_n)] = 0.
\label{eq:conn_bounding}
\end{align}
Our approach will be to find a suitable upper bound for (\ref{eq:conn_bounding}) and prove that it goes to zero as $n$ goes to infinity with $c>1$.

We now work towards deriving an upper bound for (\ref{eq:conn_bounding}); then in Section \ref{app:establishing}
we will show that the bound goes to zero as $n$ gets large.
Define the event $E_n(\pmb{\mu},\pmb{\theta},\pmb{X})$ via
\begin{equation*}
E_n(\pmb{\mu},\pmb{\theta},\pmb{X}) :=\cup_{S \subseteq \mathcal{N} : |S| \geq 1} \left[ | \cup_{i \in S} \Sigma_i | \leq X_{|S|} \right]
\end{equation*}
where $\mathcal{N}=\{1,\ldots,n\}$
and $\pmb{X} = [X_1~ \cdots~ X_n]$ is an $n$-dimensional array of integers. Let
\begin{equation}
L_n := \min\left( \left \lfloor{\frac{P}{K_1}}\right \rfloor, \left \lfloor{\frac{n}{2}}\right \rfloor \right)
\label{eq:conn_newrange_Ln}
\end{equation}
and
\begin{equation}
X_\ell=
\begin{cases} 
      \hfill \left \lfloor{\beta \ell K_1}\right \rfloor    \hfill & \ell=1,\ldots,L_n \\
      \hfill \left \lfloor{\gamma P}\right \rfloor \hfill & \ell=L_n+1,\ldots,n \\
\end{cases}
\label{eq:conn_X}
\end{equation}
for some $\beta$ and $\gamma$ in $(0,\frac{1}{2})$ that will be specified later. In words, $E_n(\pmb{\mu},\pmb{\theta},\pmb{X})$ denotes the event that  there exists $\ell=1,\ldots, n$ such that the number of unique keys stored by at least one subset of $\ell$ sensors is less than $\left \lfloor{\beta \ell K_1}\right \rfloor \pmb{1}[\ell \leq L_n] + \left \lfloor{\gamma P}\right \rfloor \pmb{1}[\ell > L_n]$. Using a crude bound, we get
\begin{align}
&\mathbb{P}[C_n(\pmb{\mu},\pmb{\Theta}_n)^c \cap I_n(\pmb{\mu},\pmb{\Theta}_n)]  \leq \mathbb{P}[E_n(\pmb{\mu},\pmb{\theta}_n,\pmb{X}_n)] + \mathbb{P}[C_n(\pmb{\mu},\pmb{\Theta}_n)^c \cap I_n(\pmb{\mu},\pmb{\Theta}_n) \cap E_n(\pmb{\mu},\pmb{\theta}_n,\pmb{X}_n)^c] \label{eq:conn_2}
\end{align}
Thus, (\ref{eq:conn_bounding}) will be established by showing that
\begin{equation}
\lim_{n \to \infty} \mathbb{P}[E_n(\pmb{\mu},\pmb{\theta}_n,\pmb{X}_n)]=0, \label{eq:conn_3} \\
\end{equation}
and
\begin{equation}
\lim_{n \to \infty} \mathbb{P}[C_n(\pmb{\mu},\pmb{\Theta}_n)^c \cap I_n(\pmb{\mu},\pmb{\Theta}_n) \cap E_n(\pmb{\mu},\pmb{\theta}_n,\pmb{X}_n)^c] =0 \label{eq:conn_4}
\end{equation}
The next proposition establishes (\ref{eq:conn_3}).
{
\prop
\label{prop:En}
Consider scalings $K_1,\ldots,K_r,P: \mathbb{N}_0 \rightarrow \mathbb{N}_0^{r+1}$ such that (\ref{scaling_condition_KG}) 
holds for some $c>1$, (\ref{eq:conn_K1}) , and (\ref{eq:conn_Pn2}) hold. Then, we have (\ref{eq:conn_3}) where $\pmb{X}_n$ is as specified in (\ref{eq:conn_X}), $\beta \in (0,\frac{1}{2})$ and $\gamma \in (0,\frac{1}{2})$ are selected such that
\begin{align}
&\max  \left( 2 \beta \sigma, \beta \left( \frac{e^2}{\sigma} \right)^{\frac{\beta}{1-2 \beta}} \right) < 1
\label{eq:conn_beta}
\\
&\max  \left( 2 \left( \sqrt{\gamma} \left( \frac{e}{\gamma} \right)^\gamma \right)^\sigma, \sqrt{\gamma} \left( \frac{e}{\gamma} \right)^\gamma \right) < 1
\label{eq:conn_gamma}
\end{align}
}

\begin{proof}
The proof is similar to \cite[Proposition~7.2]{Yagan/Inhomogeneous}. Results only require the conditions (\ref{eq:conn_Pn2}) and (\ref{eq:conn_K1_new_trick_2}) to hold. The latter condition is clearly established in Lemma~\ref{cor:new_corr}.
\end{proof}

The rest of the paper is devoted to establishing (\ref{eq:conn_4}) under the enforced assumptions on the scalings and 
with
$\pmb{X}_n$ as specified in (\ref{eq:conn_X}), $\beta \in (0,\frac{1}{2})$ selected small enough such that (\ref{eq:conn_beta}) holds, and $\gamma \in (0,\frac{1}{2})$ selected small enough such that (\ref{eq:conn_gamma}) holds.
We denote by $\mathbb{H}(n,\pmb{\mu},\pmb{\Theta}_n)(S)$ a subgraph of $\mathbb{H}(n,\pmb{\mu},\pmb{\Theta}_n)$ whose vertices are restricted to the set $S$. Define the events

\begin{align} \nonumber
 C_n(\pmb{\mu},\pmb{\Theta}_n,S) &:= [\mathbb{H}(n,\pmb{\mu},\pmb{\Theta}_n)(S) \text{ is connected}] 
\\
B_n(\pmb{\mu},\pmb{\Theta}_n,S) 
&:= [\mathbb{H}(n,\pmb{\mu},\pmb{\Theta}_n)(S) \text{ is isolated}]  \nonumber
\\
A_n(\pmb{\mu},\pmb{\Theta}_n,S) &:= C_n(\pmb{\mu},\pmb{\Theta}_n,S) \cap B_n(\pmb{\mu},\pmb{\Theta}_n,S)
\nonumber
\end{align}
In other words, $A_n(\pmb{\mu},\pmb{\Theta}_n,S)$ encodes the event that $\mathbb{H}(n,\pmb{\mu},\pmb{\Theta}_n)(S)$ is a \textit{component}, i.e., a connected subgraph that is isolated from the rest of the graph. The key observation is that a graph is {\em not} connected if and only if it has
a component on vertices $S$ with $1 \leq |S| \leq \left \lfloor{\frac{n}{2}}\right \rfloor$; note that if vertices $S$ form a component then so do vertices $\mathcal{N}-S$.
The event $I_n(\pmb{\mu},\pmb{\Theta}_n)$ eliminates the possibility of $\mathbb{H}(n,\pmb{\mu},\pmb{\Theta}_n)(S)$ containing a component of size one (i.e., an isolated node), whence we have
\begin{equation*}
C_n(\pmb{\mu},\pmb{\Theta}_n)^c \cap I_n(\pmb{\mu},\pmb{\Theta}_n) \subseteq \cup_{S \in \mathcal{N}:2 \leq |S| \leq \left \lfloor{\frac{n}{2}}\right \rfloor} A_n(\pmb{\mu},\pmb{\Theta}_n,S)
\end{equation*}
and the conclusion
\begin{equation}
\mathbb{P}[C_n(\pmb{\mu},\pmb{\Theta}_n)^c \cap I_n(\pmb{\mu},\pmb{\Theta}_n)] \leq \sum_{S \in \mathcal{N}:2 \leq |S| \leq \left \lfloor{\frac{n}{2}}\right \rfloor} \mathbb{P}[A_n(\pmb{\mu},\pmb{\Theta}_n,S)]
\nonumber
\end{equation}
follows. 
 
By exchangeability, we get
\begin{align}
\mathbb{P}[C_n(\pmb{\mu},\pmb{\Theta}_n)^c \cap I_n(\pmb{\mu},\pmb{\Theta}_n) \cap E_n(\pmb{\mu},\pmb{\theta}_n,\pmb{X}_n)^c] &\leq \sum_{\ell=2}^{\left \lfloor{\frac{n}{2}}\right \rfloor} \left( \sum_{S \in \mathcal{N}_{n,\ell}} \mathbb{P}[A_n(\pmb{\mu},\pmb{\Theta}_n,S) \cap E_n(\pmb{\mu},\pmb{\theta}_n,\pmb{X}_n)^c] \right)
\nonumber \\
&= \sum_{\ell=2}^{\left \lfloor{\frac{n}{2}}\right \rfloor} \dbinom{n}{\ell} \mathbb{P}[A_{n,\ell}(\pmb{\mu},\pmb{\Theta}_n) \cap E_n(\pmb{\mu},\pmb{\theta}_n,\pmb{X}_n)^c]
\label{eq:key_bound1_connectivity_osy}
\end{align}
where $ \mathcal{N}_{n,\ell}$ denotes the collection of all subsets of $\{1,\ldots, n\}$ with exactly $\ell$ elements,
and $A_{n,\ell}(\pmb{\mu},\pmb{\Theta}_n)$ denotes the event that the set $\{1,\ldots, \ell \}$ of nodes form a component. 
As before we have $A_{n,\ell}(\pmb{\mu},\pmb{\Theta}_n) = C_\ell(\pmb{\mu},\pmb{\Theta}_n) \cap B_{n,\ell}(\pmb{\mu},\pmb{\Theta}_n) $,
where 
$C_\ell(\pmb{\mu},\pmb{\Theta}_n)$  denotes the event that the set $\{1,\ldots,\ell\}$ of nodes is connected and
$B_{n,\ell}(\pmb{\mu},\pmb{\Theta}_n)$ denotes the event that the set $\{1,\ldots,\ell\}$ of nodes is isolated from the rest of the graph. 

It is now clear that (\ref{eq:conn_4}) is established once we show that
\begin{equation}
\lim_{n \to \infty} \sum_{\ell=2}^{\left \lfloor{\frac{n}{2}}\right \rfloor} \binom{n}{\ell} \mathbb{P}[A_{n,\ell}(\pmb{\mu},\pmb{\Theta}_n) \cap E_n(\pmb{\mu},\pmb{\theta}_n,\pmb{X}_n)^c=0.
\label{eq:to_be_established}
\end{equation} 
We proceed by deriving bounds on the probabilities appearing in (\ref{eq:to_be_established}). Conditioning on $\Sigma_1, \ldots, \Sigma_{\ell}$ and $\{B_{ij}(\pmb{\alpha}), 1 \leq i < j \leq \ell \}$, we get
\begin{align}
&\mathbb{P}\left[A_{n,\ell}(\pmb{\mu},\pmb{\Theta}_n) \cap E_n(\pmb{\mu},\pmb{\theta}_n,\pmb{X}_n)^c \right] \nonumber \\
&= \mathbb{E} \bigg[ \mathbb{E}\bigg[ \pmb{1} \left[ C_\ell \left(\pmb{\mu},\pmb{\Theta}_n \right) \cap B_{n,\ell}\left(\pmb{\mu},\pmb{\Theta}_n \right) \cap  E_n(\pmb{\mu},\pmb{\theta}_n,\pmb{X}_n)^c\right]\given[\bigg] 
\begin{array}{c}
\Sigma_1,\ldots,\Sigma_\ell \\
B_{ij}(\pmb{\alpha}),~i,j=1,\ldots,\ell
\end{array}
\bigg] \bigg] \nonumber \\
&=\mathbb{E} \Big[ \pmb{1} \left[ C_\ell \left(\pmb{\mu},\pmb{\Theta}_n \right) \right]  \cdot \mathbb{P} \Big[B_{n,\ell}\left(\pmb{\mu},\pmb{\Theta}_n \right) \cap  E_n(\pmb{\mu},\pmb{\theta}_n,\pmb{X}_n)^c  \given[\Big] 
\begin{array}{c}
\Sigma_1,\ldots,\Sigma_\ell 
\end{array}
\Big] \Big]
\label{eq:breaking_down_Anl}
\end{align}
since  $C_\ell(\pmb{\mu},\pmb{\Theta}_n)$ is 
fully determined by $\Sigma_1, \ldots, \Sigma_{\ell}$ and $\{B_{ij}(\alpha_n), 1 \leq i < j \leq \ell \}$,
and $B_{n,\ell}(\pmb{\mu},\pmb{\Theta}_n)$ and $E_n(\pmb{\mu},\pmb{\theta}_n,\pmb{X}_n)$
are independent from $\{B_{ij}(\pmb{\alpha}), 1 \leq i, j \leq \ell \}$.

Next, we consider the probabilities appearing in (\ref{eq:breaking_down_Anl}). For each $\ell=1,\ldots,n-1$, we have
\begin{equation*}
B_{n,\ell}(\pmb{\mu},\pmb{\Theta}_n) = \bigcap_{k=\ell+1}^{n} \left[\left| \cup_{i \in \nu_{\ell,k}(\pmb{\alpha})} \Sigma_i \right| \cap \Sigma_k = \emptyset  \right]
\end{equation*}
with $\nu_{\ell,k}(\pmb{\alpha})$ as defined in (\ref{eq:isolated_v(alpha)}). We have
\begin{align}
\bP{B_{n,\ell}(\pmb{\mu},\pmb{\Theta}_n) ~\big|~ \Sigma_1, \ldots, \Sigma_{\ell} } &= \mathbb{E} \Bigg[ \mathbb{E}\left[ \pmb{1}\left[B_{n,\ell}(\pmb{\mu},\pmb{\Theta}_n) \right] \given[\Bigg] \hspace{-4mm}
\begin{array}{c}
\Sigma_1, \ldots, \Sigma_n, \\
B_{ij}(\pmb{\alpha}): i=1,\ldots,\ell,\\
\quad \quad \quad \quad j=\ell+1,\ldots,n
\end{array}
\right]  \given[\Bigg]  \Sigma_1, \ldots, \Sigma_{\ell}  \Bigg] \nonumber \\
&= \mathbb{E} \left[ \mathlarger{\mathlarger{\prod}}_{k=\ell+1}^n  \frac{\dbinom{P-| \cup_{i \in \nu_{\ell,k}(\pmb{\alpha})} \Sigma_i |}{|\Sigma_k|}}{\dbinom {P}{|\Sigma_k|}} ~\bigg|~ \Sigma_1, \ldots, \Sigma_{\ell} \right] \nonumber
\end{align}
Observe that on the event $E_n(\pmb{\mu},\pmb{\theta}_n,\pmb{X}_n)^c$ we have
\begin{align}
\left| \cup_{i \in \nu_{\ell,k}(\pmb{\alpha})} \Sigma_i \right|   \geq \left( X_{n,|\nu_{\ell,k}(\pmb{\alpha})|} +1 \right) \pmb{1}[\left|\nu_{\ell,k}(\pmb{\alpha})\right| >0] \nonumber
\end{align}
Moreover, the crude bound 
\begin{align}
\left| \cup_{i \in \nu_{\ell,k}(\pmb{\alpha})} \Sigma_i \right|  \geq K_{t_{\min,\ell}} \pmb{1}[\left|\nu_{\ell,k}(\pmb{\alpha})\right| >0] \nonumber
\end{align}
always holds with $t_{\min,\ell}=\min \{t_1,\ldots,t_\ell\}$. Hence, we can write 
\begin{align}
&\mathbb{P} \left[B_{n,\ell}(\pmb{\mu},\pmb{\Theta}_n) \cap E_n(\pmb{\mu},\pmb{\theta}_n,\pmb{X}_n)^c \given[\big] \Sigma_1, \ldots, \Sigma_{\ell} \right] \nonumber \\
&\quad \leq  \mathbb{E} \vast[ \mathlarger{\mathlarger{\prod}}_{k=\ell+1}^n \frac{\dbinom{P-\max\left(K_{t_{\min,\ell}} ,  X_{n,|\nu_{\ell,k}(\pmb{\alpha})|} +1  \right)\pmb{1}[\left|\nu_{\ell,k}(\pmb{\alpha})\right| >0]}{|\Sigma_k|}}{\dbinom {P}{|\Sigma_k|}} \given[\vast]
\begin{array}{c}
\Sigma_1, \ldots, \Sigma_{\ell}
\end{array} \vast] \nonumber
\end{align}
Note that conditioned on $\Sigma_1, \Sigma_2, \ldots,\Sigma_\ell$, we can determine the class of each node in $\{1,\ldots,\ell\}$, i.e., $t_i=1 \cdot \pmb{1}\left[|\Sigma_i|=K_1\right]+2 \cdot \pmb{1}\left[|\Sigma_i|=K_2\right]+\ldots+r \cdot \pmb{1}\left[|\Sigma_i|=K_r\right] $ for $i=1,\ldots,\ell$. Moreover, since $\left| \nu_{\ell,k}(\pmb{\alpha}) \right| = \pmb{1} \left[v_1 \sim_G v_k \right] +  \pmb{1} \left[v_2 \sim_G v_k \right] + \ldots + \pmb{1} \left[v_\ell \sim_G v_k \right]$, the random variables $\{\left| \nu_{\ell,k}(\pmb{\alpha}) \right|\}_{k=\ell+1}^n$ are independent and identically distributed. In particular
\begin{equation}\nonumber
\left| \nu_{\ell,k}(\pmb{\alpha}) \right| \given[\Big]  \Sigma_1, \ldots, \Sigma_{\ell} \sim \text{Poisson-Binomial}\left(\ell,\pmb{p}=\left( \alpha_{t_1 j},  \alpha_{t_2 j}, \ldots,  \alpha_{t_\ell j} \right) \right) \quad \text{with probability } \mu_j
\end{equation}
for $k=\ell+1,4,\ldots,n$. It follows that
\begin{align}
&\mathbb{P} \left[B_{n,\ell}(\pmb{\mu},\pmb{\Theta}_n) \cap E_n(\pmb{\mu},\pmb{\theta}_n,\pmb{X}_n)^c \given[\big] \Sigma_1, \ldots, \Sigma_{\ell} \right] \nonumber \\
&\leq \left( \mathbb{E} \vast[  \frac{\dbinom{P-\max\left(K_{t_{\min,\ell}} ,  X_{n,|\nu_{\ell,\ell+1}(\pmb{\alpha})|} +1  \right)\pmb{1}[\left|\nu_{\ell,\ell+1}(\pmb{\alpha})\right| >0]}{|\Sigma_k|}}{\dbinom {P}{|\Sigma_k|}} \given[\vast]
\begin{array}{c}
\Sigma_1, \ldots, \Sigma_{\ell}
\end{array} \vast] \right)^{n-\ell} \nonumber \\
 &= \vast( \mathlarger{\mathlarger{\sum}}_{j=1}^r \mu_j \mathbb{E} \vast[ \frac{\dbinom{P-\max\left(K_{t_{\min,\ell}} ,  X_{n,|\nu_{\ell,\ell+1}(\pmb{\alpha})|} +1  \right)\pmb{1}[\left|\nu_{\ell,\ell+1}(\pmb{\alpha})\right| >0]}{K_j}}{\dbinom {P}{K_j}} \given[\vast]
\begin{array}{c}
\Sigma_1, \ldots, \Sigma_{\ell},\\
t_{\ell+1}=j
\end{array} \vast] \vast)^{n-\ell} \label{eq:bound_on_B_cap_Ec}
\end{align}
by the law of total expectation. Reporting (\ref{eq:bound_on_B_cap_Ec}) into (\ref{eq:breaking_down_Anl}), we then get

\begin{align}
&\mathbb{P}\left[A_{n,\ell}(\pmb{\mu},\pmb{\Theta}_n) \cap E_n(\pmb{\mu},\pmb{\theta}_n,\pmb{X}_n)^c \right] \leq \mathbb{E} \vast[ \pmb{1} \left[ C_\ell \left(\pmb{\mu},\pmb{\Theta}_n \right) \right]  \cdot \nonumber \\
&\quad \cdot \left( \mathlarger{\mathlarger{\sum}}_{j=1}^r \mu_j \mathbb{E} \left[ \frac{\dbinom{P-\max\left(K_{t_{\min,\ell}} ,  X_{n,|\nu_{\ell,\ell+1}(\pmb{\alpha})|} +1  \right)\pmb{1}[\left|\nu_{\ell,\ell+1}(\pmb{\alpha})\right| >0]}{K_j}}{\dbinom {P}{K_j}} \given[\vast]
\begin{array}{c}
\Sigma_1, \ldots, \Sigma_{\ell},\\
t_{\ell+1}=j
\end{array} \right] \right)^{n-\ell} \vast]
\label{eq:P_Anl_Ec_combined}
\end{align}

The following lemma gives bounds on the terms appearing in (\ref{eq:P_Anl_Ec_combined}). The proof is given in Appendix~\ref{app:proof_of_bounding_lemma}.
{\lemma
\label{lemma:final_bounds_on_Anl_Ec}
Consider a probability distribution $\pmb{\mu}=(\mu_1,\mu_2,\ldots,\mu_r)$, integers 
$K_1  \leq \cdots  \leq K_r \leq P/2$, and  $\pmb{\alpha}=\left\{ \alpha_{ij}\right\}$ for $i,j=1,\ldots,r$ with $\alpha_{ij} \in (0,1)$. With $\pmb{X}_n$ as specified in (\ref{eq:conn_X}), $\beta \in (0,\frac{1}{2})$ and $\gamma \in (0,\frac{1}{2})$ as specified in (\ref{eq:conn_beta}) and (\ref{eq:conn_gamma}) respectively, we have
\begin{align}
\mathbb{P}[C_\ell(\pmb{\mu},\pmb{\Theta})] \leq \min \left\{ 1,\ell^{\ell-2} \left( \max_{i,j} \left\{\alpha_{ij} p_{ij}\right\} \right)^{\ell-1} \right\}  
\label{eq:conn_key_bound_aux1}
\end{align}
and
\begin{align}
& \left( \mathlarger{\mathlarger{\sum}}_{j=1}^r \mu_j \mathbb{E} \left[ \frac{\dbinom{P-\max\left( K_{t_{\min,\ell}} ,  X_{n,|\nu_{\ell,\ell+1}(\pmb{\alpha})|} +1  \right)\pmb{1}[\left|\nu_{\ell,\ell+1}(\pmb{\alpha})\right| >0]}{K_j}}{\dbinom {P}{K_j}}  \given[\vast]
\begin{array}{c}
\Sigma_1, \ldots, \Sigma_{\ell},\\
t_{\ell+1}=j
\end{array} \right] \right)^{n-\ell} \nonumber \\
&\leq  \left( \min \left\{1-\Lambda_m, \min \left\{1-\mu_r + \mu_r e^{-\alpha_{\min} p_{1r} \beta \ell}, e^{-\alpha_{\min} p_{11} \beta \ell} \right\} + e^{- \gamma K_{1}} \pmb{1}\left[\ell>L_n\right]\right\} \right)^{n-\ell}
\label{eq:conn_key_bound_aux2}
\end{align}
}

Note that as we report (\ref{eq:conn_key_bound_aux2}) back into (\ref{eq:P_Anl_Ec_combined}), we get
\begin{align}
&\mathbb{P}\left[A_{n,\ell}(\pmb{\mu},\pmb{\Theta}_n) \cap E_n(\pmb{\mu},\pmb{\theta}_n,\pmb{X}_n)^c \right] \nonumber \\
&\leq \mathbb{E} \Bigg[ \pmb{1} \left[ C_\ell \left(\pmb{\mu},\pmb{\Theta}_n \right) \right]  \cdot \nonumber \\
& \qquad \cdot \left( \min \left\{1-\Lambda_m, \min \left\{1-\mu_r + \mu_r e^{-\alpha_{\min} p_{1r} \beta \ell}, e^{-\alpha_{\min} p_{11} \beta \ell} \right\} + e^{- \gamma K_{1}} \pmb{1}\left[\ell>L_n\right]\right\} \right)^{n-\ell} \Bigg] \nonumber \\
& =  \mathbb{P}[C_\ell(\pmb{\mu},\pmb{\Theta})]  \cdot \left( \min \left\{1-\Lambda_m, \min \left\{1-\mu_r + \mu_r e^{-\alpha_{\min} p_{1r} \beta \ell}, e^{-\alpha_{\min} p_{11} \beta \ell} \right\} + e^{- \gamma K_{1}} \pmb{1}\left[\ell>L_n\right]\right\} \right)^{n-\ell}
\label{eq:P_Anl_Ec_combined_Final}
\end{align}

In addition, it holds that
\begin{equation}
\max_{i,j} \left\{\alpha_{ij} p_{ij}\right\} \leq \alpha_{\max} p_{rr} 
\label{eq:eletreby_boundOnLargestEdgeProb}
\end{equation}

Our proof of  (\ref{eq:conn_4}) will be completed (see (\ref{eq:key_bound1_connectivity_osy})) upon establishing 
\begin{equation}
\lim_{n \to \infty} \sum_{\ell=2}^{\left \lfloor{\frac{n}{2}}\right \rfloor} \dbinom{n}{\ell} \mathbb{P}[A_{n,\ell}(\pmb{\mu},\pmb{\Theta}_n) \cap E_n(\pmb{\mu},\pmb{\theta}_n,\pmb{X}_n)^c]=0
\label{eq:conn_key}
\end{equation}
by means of (\ref{eq:conn_key_bound_aux1}), (\ref{eq:conn_key_bound_aux2}), and (\ref{eq:P_Anl_Ec_combined_Final}).
These steps are taken in the next section.

\section{Establishing (\ref{eq:conn_key})}
\label{app:establishing}
We will establish (\ref{eq:conn_key}) in several steps with each step focusing on a specific range of the summation over $\ell$. Throughout, we consider a scalings $K_1,\ldots,K_r,P: \mathbb{N}_0 \rightarrow \mathbb{N}_0^{r+1}$ and $\pmb{\alpha}: \mathbb{N}_0 \rightarrow (0,1)^{r \times r}$ such that (\ref{scaling_condition_KG}) 
holds with $c>1$, (\ref{eq:conn_K1}), (\ref{eq:conn_K1r}), (\ref{eq:bounding_variance_of_alpha}), and (\ref{eq:conn_Pn2}) hold.

\subsection{The case where $2 \leq \ell \leq R$}
This range considers fixed values of $\ell$. Pick an integer $R$ to be specified later at (\ref{eq:choosing_R}). Use (\ref{scaling_condition_KG}), (\ref{eq:equation_bounding_cl_event}), (\ref{eq:isolated:exp_bound}), (\ref{eq:conn_bounds_1}), (\ref{eq:conn_key_bound_aux1}), the first bound in (\ref{eq:conn_key_bound_aux2}), and (\ref{eq:eletreby_boundOnLargestEdgeProb}) to get
\begin{align*}
\sum_{\ell=2}^{R} \dbinom{n}{\ell} \mathbb{P}[A_{n,\ell}(\pmb{\mu},\pmb{\Theta}_n) \cap E_n(\pmb{\mu},\pmb{\theta}_n,\pmb{X}_n)^c] &\leq \sum_{\ell=2}^{R} \left( \frac{en}{\ell}\right)^\ell  \ell^{\ell-2} \left( \alpha_{\max}(n) p_{rr}(n)\right)^{\ell-1} \left( 1- \Lambda_m(n) \right)^{n-\ell}\nonumber \\
& \leq \sum_{\ell=2}^{R} \left( en \right)^\ell  \left( \frac{(\log n)^{\tau+2}}{n}\right)^{\ell-1} \left( 1-c_n \frac{\log n}{n}\right)^{n-\ell}\nonumber \\
&\leq \sum_{\ell=2}^{R}  n \left( e(\log n)^{\tau+2}\right)^{\ell} e^{-c_n \log n \frac{n-\ell}{n}}\nonumber \\
&=\sum_{\ell=2}^{R} \left( e(\log n)^{\tau+2}\right)^{\ell}  n^{1-c_n\frac{n-\ell}{n}} \nonumber
\end{align*}
With $c>1$, we have $\lim_{n \to \infty} \left(1-c_n\frac{n-\ell}{n}\right)=1-c<0$.
Thus, 
for each $\ell=2,3, \ldots, R$ and a finite $\tau>0$, we have
\begin{equation} \nonumber
\left( e(\log n)^{\tau+2}\right)^{\ell-1}  n^{1-c_n\frac{n-\ell}{n}}=o(1),
\end{equation}
whence we get
\begin{equation} \nonumber
\lim_{n \to \infty} \sum_{\ell=2}^{R} \dbinom{n}{\ell} \mathbb{P}[A_{n,\ell}(\pmb{\mu},\pmb{\Theta}_n) \cap E_n(\pmb{\mu},\pmb{\theta}_n,\pmb{X}_n)^c] =0.
\end{equation}

\subsection{The case where $R+1 \leq \ell \leq \min\{L_n,  \lfloor{\frac{\mu_r n}{\beta c_n \log n}} \rfloor \}$}
\label{subsection:geometric_series}
Our goal in this and the next subsection is to cover the range $R+1 \leq \ell \leq  \lfloor{\frac{\mu_r n}{\beta c_n \log n}} \rfloor $.
Since the bound given at (\ref{eq:conn_key_bound_aux2}) takes a different form when $\ell > L_n$, we first consider the 
range $R+1 \leq \ell \leq \min\{L_n,  \lfloor{\frac{\mu_r n}{\beta c_n \log n}} \rfloor \}$.
Using (\ref{eq:equation_bounding_cl_event}), (\ref{eq:isolated:exp_bound}), (\ref{eq:conn_bounds_1}), (\ref{eq:conn_key_bound_aux1}), the second bound in (\ref{eq:conn_key_bound_aux2}), and (\ref{eq:eletreby_boundOnLargestEdgeProb}) we get
\begin{align}
&\sum_{\ell=R+1}^{\min\{L_n, \lfloor{\frac{\mu_r n}{\beta c_n \log n}} \rfloor\}} \dbinom{n}{\ell} \mathbb{P}[A_{n,\ell}(\pmb{\mu},\pmb{\Theta}_n) \cap E_n(\pmb{\mu},\pmb{\theta}_n,\pmb{X}_n)^c] \nonumber \\
& \leq \sum_{\ell=R+1}^{\min\{L_n, \lfloor{\frac{\mu_r n}{\beta c_n \log n}} \rfloor \}}  \left( \frac{en}{\ell}\right)^\ell  \ell^{\ell-2} 
\left( \frac{(\log n)^{\tau+2}}{n}\right)^{\ell-1}  \cdot \Bigg( 1 -\mu_r \left( 1-e^{- \alpha_{\min}(n) \beta \ell p_{1r}(n)} \right) \Bigg)^{n-\ell}
 \label{eq:conn_11}
\end{align}
From the upper bound in (\ref{eq:bound_on_alpha_min_p1r_part1}) and $\ell \leq \frac{\mu_r n}{ \beta c_n \log n}$, we have 
\begin{align*}
\alpha_{\min}(n) \beta \ell p_{1r}(n)  &\leq  \beta \frac{\mu_r n}{ \beta c_n \log n} \frac{c_n}{\mu_r} \frac{\log n}{n}  =  1.
\end{align*}
Using the fact that
$
1-e^{-x} \geq \frac{x}{2}$ for all $0 \leq x \leq 1$,
we get
\begin{align}
1-\mu_r \hspace{-.7mm}\left( 1-e^{- \alpha_{\min}(n) \beta \ell p_{1r}(n)} \right) \hspace{-.7mm} &\leq 1  -  \frac{\mu_r \alpha_{\min}(n) \beta \ell p_{1r}(n)}{2}  \leq \hspace{-.5mm} e^{- \beta \ell \mu_r \rho \frac{\log n}{2n}}
\label{eq:conn_12}
\end{align}
using the lower bound in (\ref{eq:conn_K1r_modified}).
Reporting this last bound in to (\ref{eq:conn_11}) and noting that 
\begin{equation}
n-\ell \geq \frac{n}{2}, \qquad \ell = 2, 3, \ldots, \left \lfloor \frac{n}{2} \right \rfloor,
\label{eq:bounding_n_minus_l}
\end{equation}
we get
\begin{align}
\sum_{\ell=R+1}^{\min\{L_n,\left \lfloor{\frac{\mu_r n}{\beta c_n \log n}}\right \rfloor\}} \dbinom{n}{\ell} \mathbb{P}[A_{n,\ell}(\pmb{\mu},\pmb{\Theta}_n) \cap E_n(\pmb{\mu},\pmb{\theta}_n,\pmb{X}_n)^c]& \leq \sum_{\ell=R+1}^{\min\{L_n, \lfloor{\frac{\mu_r n}{\beta c_n \log n}} \rfloor\}} n \left( e(\log n)^{\tau+2}\right)^{\ell} 
e^{- \beta \ell \mu_r \rho \frac{\log n}{2n} \frac{n}{2}}\nonumber \\
&\leq n \sum_{\ell=R+1}^{\min\{L_n, \lfloor{\frac{\mu_r n}{\beta c_n \log n}} \rfloor\}} \left( e \left( \log n\right)^{\tau+2} e^{- \beta \rho \frac{\mu_r}{4} \log n} \right)^\ell\nonumber \\
&\leq n \sum_{\ell=R+1}^{\infty} \left( e \left( \log n\right)^{\tau+2} e^{- \beta \rho  \frac{\mu_r}{4} \log n} \right)^\ell
\label{eq:conn_13}
\end{align}

Given that $\beta, \rho, \mu_r>0$ and $\tau$ is finite, we clearly have
\begin{equation}
 e \left( \log n\right)^{\tau+2} e^{- \beta \rho \log n \frac{\mu_r}{4}} = o(1).
 \label{eq:summable_sequence}
 \end{equation}
Thus, the geometric series in (\ref{eq:conn_13}) is summable for $n$ sufficiently large, and we have
\begin{align*}
\sum_{\ell=R+1}^{\min\{L_n, \lfloor{\frac{\mu_r n}{\beta c_n \log n}} \rfloor\}} \dbinom{n}{\ell} \mathbb{P}[A_{n,\ell}(\pmb{\mu},\pmb{\Theta}_n) \cap E_n(\pmb{\mu},\pmb{\theta}_n,\pmb{X}_n)^c] & \leq \left(1+o(1)\right) n \left( e \left( \log n\right)^{\tau+2} e^{- \beta \rho \log n \frac{\mu_r}{4}} \right)^{R+1}\nonumber \\
&=\left(1+o(1)\right) n^{1-(R+1){\beta \rho}\frac{\mu_r}{4}} \left( e (\log n)^{\tau+2} \right)^{R+1} \nonumber \\
& = o(1)
\end{align*}
for any positive integer $R$ with
\begin{equation} 
R>\frac{8}{\beta \rho \mu_r}. 
\label{eq:choosing_R}
\end{equation}
This choice is permissible given that $\rho, \beta, \mu_r > 0$. 

\subsection{The case where $ \min \{ \lfloor{\frac{\mu_r n}{ \beta c_n \log n}} \rfloor,\max (R,L_n) \} < \ell \leq  \lfloor{\frac{\mu_r n}{ \beta c_n \log n}} \rfloor$}

Clearly, this range becomes obsolete if $\max (R,L_n) \geq  \lfloor{\frac{\mu_r n}{ \beta c_n \log n}} \rfloor$. 
Thus, it suffices to consider the subsequences for which the range $\max (R,L_n)+1 \leq \ell \leq  \lfloor{\frac{\mu_r n}{\beta c_n \log n}} \rfloor$ is non-empty. There, we use (\ref{eq:equation_bounding_cl_event}), (\ref{eq:isolated:exp_bound}), (\ref{eq:conn_bounds_1}), (\ref{eq:conn_key_bound_aux1}), the second bound in (\ref{eq:conn_key_bound_aux2}), and (\ref{eq:eletreby_boundOnLargestEdgeProb}) to get
\begin{align}
&\sum_{\ell=\max (R,L_n)+1}^{\left \lfloor{\frac{\mu_r n}{\beta c_n \log n}}\right \rfloor} \dbinom{n}{\ell}  \mathbb{P}[A_{n,\ell}(\pmb{\mu},\pmb{\Theta}_n) \cap E_n(\pmb{\mu},\pmb{\theta}_n,\pmb{X}_n)^c] \label{eq:conn_newrange_2}
 \\
&\leq \sum_{\ell=\max (R,L_n)+1}^{\left \lfloor{\frac{\mu_r n}{ \beta c_n \log n}}\right \rfloor} \left( \frac{en}{\ell} \right)^\ell \ell^{\ell-2} \left( \frac{\left(\log n \right)^{\tau+2}}{n} \right)^{\ell-1} \cdot \left( 1-\mu_r \left(1-e^{-\beta \ell \alpha_{\min}(n) p_{1r}(n)} \right)+e^{-\gamma K_{1,n}} \right)^{\frac{n}{2}}
\nonumber \\
& \leq \hspace{-2mm} \sum_{\ell=\max (R,L_n)+1}^{\left \lfloor{\frac{\mu_r n}{2 \beta c \log n}}\right \rfloor} \hspace{-3mm} n \left(e \left(\log n \right)^{\tau+2} \right)^\ell \left( e^{- \beta \ell \rho \mu_r \frac{\log n}{2n} } + e^{-\gamma K_{1,n}}\right)^{\frac{n}{2}}
\nonumber
\end{align}
where in the last step we used
 (\ref{eq:conn_12}) in view of $\ell \leq \frac{\mu_r n}{\beta c_n \log n}$.
 
Next, we write
\begin{align}
e^{- \beta \ell \rho \mu_r \frac{\log n}{2n} } + e^{-\gamma K_{1,n}} &=  e^{- \beta \ell \rho \mu_r \frac{\log n}{2n} } \left(1+e^{-\gamma K_{1,n} + \beta \ell \rho \mu_r \frac{\log n}{2n}} \right) \nonumber \\
& \leq \exp \left\{- \beta \ell \rho \mu_r \frac{\log n}{2n} + e^{-\gamma K_{1,n} + \beta \ell \rho \mu_r \frac{\log n}{2n}} \right\} \nonumber \\
& \leq \exp \left\{- \beta \ell \rho  \mu_r \frac{\log n}{2n}\left(1 - \frac{e^{-\gamma K_{1,n} + \frac{\rho \mu_r^2 }{2 c_n}}}{\beta \ell \rho \mu_r \frac{\log n}{2n}} \right) \right\} \label{eq:intermediary_range_new}
\end{align}
where the last inequality is obtained from $\ell \leq {\frac{\mu_r n}{ \beta c_n \log n}}$. 
Using the fact that $\ell > L_n=\min\{  \lfloor{\frac{P_n}{K_{1,n}}} \rfloor,  \lfloor{\frac{n}{2}} \rfloor \}$
 and (\ref{eq:conn_Pn2}) we have 
\begin{align}
\frac{e^{-\gamma K_{1,n}}}{\beta \ell \rho \mu_r \frac{\log n}{2n}}  & \leq 
\max\left\{  \frac{K_{1,n}}{P_n},  \frac{2}{n} \right \}{2n}
\frac{e^{-\gamma K_{1,n}}}{\beta \rho \mu_r \log n} \leq \max \left \{ \frac{2 K_{1,n} e^{-\gamma K_{1,n}}}{\beta  \rho \mu_r \sigma \log n},\frac{4 e^{-\gamma K_{1,n}}}{\beta  \rho \mu_r \log n} \right\} =o(1) \nonumber
\end{align}
by virtue of (\ref{eq:conn_K1_new_trick_2}) and  the facts that $\beta, \mu_r, \sigma, \rho >0$. Reporting this
into (\ref{eq:intermediary_range_new}), we see that for
for any $\epsilon>0$, there exists a finite integer $n^*(\epsilon)$ such that 
\begin{equation}
\left( e^{- \beta \ell \rho \mu_r \frac{\log n}{2n}} + e^{-\gamma K_{1,n}}\right) \leq e^{- \beta \ell \rho \mu_r \frac{\log n}{2n} (1-\epsilon)} 
\label{eq:conn_newrange_3}
\end{equation}
for all $n \geq n^*(\epsilon)$.
Using (\ref{eq:conn_newrange_3}) in (\ref{eq:conn_newrange_2}), we get
\begin{align}
\sum_{\ell=\max(R,L_n)+1}^{\left \lfloor{\frac{\mu_r n}{ \beta c_n \log n}}\right \rfloor} \dbinom{n}{\ell}  \mathbb{P}[A_{n,\ell}(\pmb{\mu},\pmb{\Theta}_n) \cap E_n(\pmb{\mu},\pmb{\theta}_n,\pmb{X}_n)^c] &\leq n \sum_{\ell=\max(R,L_n)+1}^{\left \lfloor{\frac{\mu_r n}{ \beta c_n \log n}}\right \rfloor} \left( e \left(\log n\right)^{\tau+2} e^{- \beta \rho \mu_r \frac{\log n}{2n} (1-\epsilon) \frac{n}{2}}  \right)^\ell \nonumber \\
&\leq n \sum_{\ell=\max (R,L_n)+1}^{\infty} \left( e \left(\log n\right)^{\tau+2} e^{- \beta \rho \mu_r \frac{\log n}{4} (1-\epsilon)}  \right)^\ell
\label{eq:conn_newrange_4}
\end{align}
Similar to (\ref{eq:summable_sequence}), we have
$ \left( e \left(\log n\right)^{\tau+2} e^{- \beta \rho \mu_r \frac{\log n}{4} (1-\epsilon)} \right) = o(1)$ so that the sum 
in (\ref{eq:conn_newrange_4}) converges. 
Following a similar approach to that in Section~\ref{subsection:geometric_series}, we then see that 
\begin{equation*}
\lim_{n \to \infty} \sum_{\ell=\max (R,L_n)+1}^{\left \lfloor{\frac{\mu_r n}{2 \beta c \log n}}\right \rfloor} \dbinom{n}{\ell} \mathbb{P}[A_{n,\ell}(\pmb{\mu},\pmb{\Theta}_n) \cap E_n(\pmb{\mu},\pmb{\theta}_n,\pmb{X}_n)^c] = 0   
\end{equation*}
with $R$ selected according to (\ref{eq:choosing_R}) and $\epsilon<1/2$.

\subsection{The case where $\lfloor{\frac{\mu_r n}{\beta c_n \log n}} \rfloor+1 \leq \ell \leq \left \lfloor{\nu n}\right \rfloor $} 
\label{subsec:range_4}
We  consider $\lfloor{\frac{\mu_r n}{ \beta c_n \log n}} \rfloor+1 \leq \ell \leq \left \lfloor{\nu n}\right \rfloor$ for some $\nu \in (0,\frac{1}{2})$ to be specified later. Recall (\ref{eq:conn_K1r_modified}), (\ref{eq:conn_bounds_1}), the first bound in (\ref{eq:conn_key_bound_aux1}), and the second bound in (\ref{eq:conn_key_bound_aux2}).
Noting that $\dbinom{n}{\ell}$ is monotone increasing in $\ell$ when $0 \leq \ell \leq \left \lfloor{\frac{n}{2}}\right \rfloor$  
 and using (\ref{eq:bounding_n_minus_l}) we get
\begin{align}
&\sum_{\ell= \lfloor{\frac{\mu_r n}{\beta c_n \log n}} \rfloor +1}^{ \left \lfloor{\nu n}\right \rfloor} \dbinom{n}{\ell} \mathbb{P}[A_{n,\ell}(\pmb{\mu},\pmb{\Theta}_n) \cap E_n(\pmb{\mu},\pmb{\theta}_n,\pmb{X}_n)^c] \nonumber \\
&\leq  \hspace{-3mm}\sum_{\ell= \lfloor{\frac{\mu_r n}{ \beta c_n \log n}} \rfloor+1}^{\left \lfloor{\nu n} \right \rfloor} \hspace{-2mm}\dbinom{n}{\left \lfloor{\nu n}\right \rfloor}  \cdot \left( 1-\mu_r+\mu_re^{- \alpha_{\min}(n) \beta \ell p_{1r}(n)} + e^{- \gamma K_{1,n}}\right)^{\frac{n}{2}}\nonumber \\
& \leq \hspace{-2mm} \sum_{\ell= \lfloor{\frac{\mu_r n}{ \beta c_n \log n}} \rfloor+1}^{\left \lfloor{\nu n}\right \rfloor} \hspace{-1mm} \left( \frac{e}{\nu} \right)^{\nu n} \cdot  \left( 1-\mu_r +\mu_re^{- \beta  \frac{\mu_r n}{ \beta c_n \log n}  \frac{\rho \log n}{ n }}+ e^{- \gamma K_{1,n}}\right)^{\frac{n}{2}} \nonumber \\
&\leq n \left( \frac{e}{\nu} \right)^{\nu n}  \left( 1-\mu_r+\mu_r e^{- \frac{\rho \mu_r}{c_n}} + e^{- \gamma K_{1,n}} \right)^{\frac{n}{2}}
\nonumber \\
&= n \left( \left( \frac{e}{\nu} \right)^{2\nu} \left( 1-\mu_r+\mu_r e^{- \frac{\rho \mu_r}{c_n}} + e^{-\gamma K_{1,n} }\right)\right)^{\frac{n}{2}}
\label{eq:conn_15}
\end{align}

We have $1-\mu_r+\mu_r e^{- \frac{\rho \mu_r}{c_n}}<1$ from $\mu_r, \rho, c>0$ and 
$e^{- \gamma K_{1,n}}=o(1)$ from
 (\ref{eq:conn_K1_new_trick_2}). Also,  it holds that $\lim_{\nu \to 0} \left( \frac{e}{\nu}\right)^{2\nu}=1$.
Thus, if we pick $\nu$ small enough to ensure that
\begin{equation}
\left( \frac{e}{\nu} \right)^{2\nu} \left( 1-\mu_r+\mu_r e^{- \frac{\rho \mu_r}{c_n}} \right) < 1,
\label{eq:conn_16}
\end{equation}
then for any $0< \epsilon < 1- \left( {e}/{\nu} \right)^{2\nu} \left( 1-\mu_r+\mu_r e^{- \frac{\rho \mu_r}{c_n}} \right)$ there exists a finite integer $n^{\star}(\epsilon)$  such that
\[
 \left( \frac{e}{\nu} \right)^{2\nu} \left( 1-\mu_r+\mu_r e^{- \frac{\rho \mu_r}{c_n}} + e^{-\gamma K_{1,n} }\right) \leq 1- \epsilon, \quad  \forall n \geq n^{\star}(\epsilon).
\]
Reporting this into (\ref{eq:conn_15}), we get
\begin{equation} \nonumber
\lim_{n \to \infty} \sum_{\ell=\left \lfloor{\frac{\mu_r n}{2 \beta c \log n}}\right \rfloor+1}^{\left \lfloor{\nu n}\right \rfloor} \dbinom{n}{\ell} \mathbb{P}[A_{n,\ell}(\pmb{\mu},\pmb{\Theta}_n) \cap E_n(\pmb{\mu},\pmb{\theta}_n,\pmb{X}_n)^c]=0
\end{equation}
since $\lim_{n \to \infty} n (1-\epsilon)^{n/2} = 0$.

\subsection{The case where $\left \lfloor{\nu n}\right \rfloor+1 \leq \ell \leq \lfloor{\frac{n}{2}} \rfloor $}
\label{subsec:range_5}
In this range, we use 
 (\ref{eq:conn_bounds_2}), the first bound in (\ref{eq:conn_key_bound_aux1}), the last bound in (\ref{eq:conn_key_bound_aux2}), and (\ref{eq:bounding_n_minus_l}) to get
 \begin{align*}
\sum_{\ell=\left \lfloor{\nu n}\right \rfloor+1}^{\left \lfloor{\frac{n}{2}}\right \rfloor} \dbinom{n}{\ell} \mathbb{P}[A_{n,\ell}(\pmb{\mu},\pmb{\Theta}_n) \cap E_n(\pmb{\mu},\pmb{\theta}_n,\pmb{X}_n)^c] &\leq \sum_{\ell=\left \lfloor{\nu n}\right \rfloor+1}^{\left \lfloor{\frac{n}{2}}\right \rfloor} \dbinom{n}{\ell} \left( e^{- \beta \ell \alpha_{\min}(n) p_{11}(n)} + e^{- \gamma K_{1,n}}\right)^{\frac{n}{2}}\nonumber \\
&\leq \left(\sum_{\ell=\left \lfloor{\nu n}\right \rfloor+1}^{\left \lfloor{\frac{n}{2}}\right \rfloor} \dbinom{n}{\ell} \right) \left( e^{- \beta \nu n \alpha_{\min}(n) p_{11}(n)} + e^{- \gamma K_{1,n}}\right)^{\frac{n}{2}} \nonumber
\\
& \leq  \left( 4 e^{- \beta \nu n \alpha_{\min}(n) p_{11}(n)} + 4 e^{- \gamma K_{1,n}}\right)^{\frac{n}{2}} 
\end{align*}
 
With $\beta,\nu,\gamma>0$ have
$e^{- \beta \nu n \alpha_{\min}(n) p_{11}(n)} = o(1)$ from (\ref{eq:conn_K1}) 
and $e^{- \gamma K_{1,n}}=o(1)$ from
 (\ref{eq:conn_K1_new_trick_2}).
The conclusion
\begin{equation} \nonumber
\lim_{n \to \infty} \sum_{\ell=\left \lfloor{\nu n}\right \rfloor+1}^{\left \lfloor{\frac{n}{2}}\right \rfloor} \dbinom{n}{\ell} \mathbb{P}[A_{n,\ell}(\pmb{\mu},\pmb{\Theta}_n) \cap E_n(\pmb{\mu},\pmb{\theta}_n,\pmb{X}_n)^c]=0
\end{equation}
immediately follows and
 the proof of  one-law is completed.
\myendpf

\bibliographystyle{IEEEtran}
\bibliography{IEEEabrv,references}

\newpage

\setcounter{equation}{0}
\renewcommand{\theequation}{\thesection.\arabic{equation}}
\numberwithin{equation}{section}

\appendix
\section{Establishing Lemma~\ref{lemma:final_bounds_on_Anl_Ec}}
\label{app:proof_of_bounding_lemma}

The bounds given at Lemma \ref{lemma:final_bounds_on_Anl_Ec}
are valid irrespective of how the parameters involved scale with $n$. 
Thus, we  consider fixed $\pmb{\Theta}$ with constraints given in the statement of Lemma \ref{lemma:final_bounds_on_Anl_Ec}.

Recall that conditioned on $\Sigma_1,\Sigma_2,\ldots,\Sigma_\ell$ and $t_{\ell+1}=j$, the rv $\left|\nu_{\ell,\ell+1}(\pmb{\alpha})\right| $ is distributed as a Poisson-Binomial rv with $\ell$ trials and success probability vector $\pmb{p}=\left\{ \alpha_{t_1j},\ldots,\alpha_{t_\ell j}\right\}$. With a slight abuse of notation, let $W_{\ell,j}=1-p_{t_{\min,\ell}j}$.
Using a crude  bound and then (\ref{eq:isolated:combinatorial_bound})   we get 
\begin{align}
&\mathbb{E} \vast[ \frac{\dbinom{P-\max \left(K_{t_{\min,\ell}},X_{n,\left|\nu_{\ell,\ell+1}(\pmb{\alpha})\right|} +1\right)\pmb{1}[\left|\nu_{\ell,\ell+1}(\pmb{\alpha})\right| >0]}{K_j}}{\dbinom {P}{K_j}} \given[\vast]
\begin{array}{c}
\Sigma_1, \ldots, \Sigma_{\ell},\\
t_{\ell+1}=j
\end{array}\vast] \nonumber \\
&\leq \mathbb{E} \left[ \frac{\dbinom{P-K_{t_{\min,\ell}}\pmb{1}[\left|\nu_{\ell,\ell+1}(\pmb{\alpha})\right| >0]}{K_j}}{\dbinom {P}{K_j}} \given[\vast]
\begin{array}{c}
\Sigma_1, \ldots, \Sigma_{\ell},\\
t_{\ell+1}=j
\end{array}\right]\nonumber \\
& \leq \mathbb{E} \left[ W_{\ell,j}^{\pmb{1}[\left|\nu_{\ell,\ell+1}(\pmb{\alpha})\right| >0]} \given[\bigg]
\begin{array}{c}
\Sigma_1, \ldots, \Sigma_{\ell},\\
t_{\ell+1}=j
\end{array}\right]\nonumber \\
&=\prod_{k=1}^\ell \left(1-\alpha_{t_k j}\right) + \left( 1- \prod_{k=1}^\ell \left(1-\alpha_{t_k j}\right)  \right) W_{\ell,j}\nonumber \\
&=\prod_{k=1}^\ell \left(1-\alpha_{t_k j}\right)  (1-W_{\ell,j}) + W_{\ell,j} \nonumber \\
&\leq \left( 1-\alpha_{t_{\min,\ell}j} \right)(1-W_{\ell,j}) + W_{\ell,j}  \nonumber \\
&= 1-\alpha_{t_{\min,\ell}j}p_{t_{\min,\ell}j}.
\label{eq:conn_crude_bound_1}
\end{align}
upon noting that $ \alpha_{t_k j} <1$ for $k=1,\ldots,\ell$ and $j=1,\ldots,r$. It is now immediate that
\begin{align}
\sum_{j=1}^r \mu_j \left(1-\alpha_{t_{\min,\ell}j}p_{t_{\min,\ell}j}\right) 
&=1-\Lambda_{t_{\min,\ell}}  \leq 1-\Lambda_m
\label{eq:eletreby_boundOnIsolatedComponent}
\end{align}


Next, consider range $\ell=1,\ldots,L_n$, where we have
\begin{equation*}
\left(X_{n,\left|\nu_{\ell,\ell+1}(\pmb{\alpha})\right|} +1\right) \pmb{1}[\left|\nu_{\ell,\ell+1}(\pmb{\alpha})\right|>0] \geq \left \lceil{\beta \left|\nu_{\ell,\ell+1}(\pmb{\alpha})\right| K_1}\right \rceil
\end{equation*}

With a slight abuse of notation, let $Z_j=1-p_{1j}$. Recalling (\ref{eq:isolated:combinatorial_bound}), we get
\begin{align}
&\mathbb{E} \vast[ \frac{\dbinom{P-\max \left(K_{t_{\min,\ell}},X_{n,\left|\nu_{\ell,\ell+1}(\pmb{\alpha})\right|} +1\right)\pmb{1}[\left|\nu_{\ell,\ell+1}(\pmb{\alpha})\right| >0]}{K_j}}{\dbinom {P}{K_j}}  \given[\vast]
\begin{array}{c}
\Sigma_1, \ldots, \Sigma_{\ell},\\
t_{\ell+1}=j
\end{array}\vast]\nonumber \\
&\leq \mathbb{E} \vast[ \frac{\dbinom{P-\left \lceil{\beta \left|\nu_{\ell,\ell+1}(\pmb{\alpha})\right| K_1}\right \rceil}{K_j}}{\dbinom {P}{K_j}}  \given[\vast]
\begin{array}{c}
\Sigma_1, \ldots, \Sigma_{\ell},\\
t_{\ell+1}=j
\end{array}\vast]\nonumber \\
&\leq \mathbb{E}\left[ Z_j^{\beta \left|\nu_{\ell,\ell+1}(\pmb{\alpha})\right|}\given[\bigg]
\begin{array}{c}
\Sigma_1, \ldots, \Sigma_{\ell},\\
t_{\ell+1}=j
\end{array}\right]
 \label{eq:conn_lemma_1}
\end{align}
Recall that 
\begin{equation} \nonumber
\left|\nu_{\ell,\ell+1}(\pmb{\alpha})\right| = \pmb{1} \left[v_1 \sim_G v_{\ell+1} \right] + \pmb{1} \left[v_2 \sim_G v_{\ell+1} \right] + \ldots +  \pmb{1} \left[v_\ell \sim_G v_{\ell+1} \right]
\end{equation}
and note that conditioned on $\Sigma_1, \ldots, \Sigma_\ell$ and that $t_{\ell+1} = j$, the indicator random variables $\pmb{1} \left[v_i \sim_G v_{\ell+1} \right]$ are each distributed as a Bernoulli random variable with parameter $\alpha_{t_i j}$ for $i=1,\ldots,r$, where $t_i$ denotes the class of node $v_i$. Let $\alpha_{\min_j} = \min \left\{\alpha_{1j}, \alpha_{2j}, \ldots, \alpha_{rj}   \right\}$. It follows that
\begin{equation} \nonumber
\left|\nu_{\ell,\ell+1}(\pmb{\alpha})\right| \succeq \left|\nu_{\ell,\ell+1}(\alpha_{\min_j})\right|
\end{equation}
where $\left|\nu_{\ell,\ell+1}(\alpha_{min_j})\right|$ denotes a binomial rv with parameters $\ell$ and $\alpha_{\min_j}$, and the operator $\succeq$ denotes the usual stochastic ordering. It follows that
\begin{align}
&\mathbb{E} \vast[ \frac{\dbinom{P-\max (K_{t_{\min,\ell}},X_{n,\left|\nu_{\ell,\ell+1}(\pmb{\alpha})\right|} +1)\pmb{1}[\left|\nu_{\ell,\ell+1}(\pmb{\alpha})\right| >0]}{K_j}}{\dbinom {P}{K_j}}  \given[\vast]
\begin{array}{c}
\Sigma_1, \ldots, \Sigma_{\ell},\\
t_{\ell+1}=j
\end{array}\vast]\nonumber \\
&\leq \mathbb{E}\left[ Z_j^{\beta \left|\nu_{\ell,\ell+1}(\pmb{\alpha})\right|}\given[\bigg]
\begin{array}{c}
\Sigma_1, \ldots, \Sigma_{\ell} \nonumber \\
t_{\ell+1}=j
\end{array}\right], \\
&\leq \mathbb{E}\left[ Z_j^{\beta \left|\nu_{\ell,\ell+1}(\alpha_{\min_j})\right|} \given[\bigg]
\begin{array}{c}
\Sigma_1, \ldots, \Sigma_{\ell},\\
t_{\ell+1}=j
\end{array}\right] \nonumber \\
&= \sum_{k=0}^\ell \dbinom {\ell}{k} \alpha_{\min_j}^k (1-\alpha_{\min_j})^{\ell-k} Z_j^{\beta k}  \nonumber \\
&=\left( 1-\alpha_{\min_j}\left(1-Z_j^\beta \right) \right)^\ell  \nonumber \\
& \leq \left( 1-\alpha_{\min_j} \beta \left(1-Z_j\right) \right)^\ell \nonumber \\
& \leq e^{-\alpha_{\min_j}(1-Z_j)\beta \ell}\nonumber \\
&=e^{-\alpha_{\min_j}p_{1j}\beta \ell}
 \label{eq:conn_lemma_1_dash}
\end{align}
using the fact 
that 
$1-Z_j^\beta \geq \beta(1-Z_j)$ 
with $Z_j \leq 1$ and $0 \leq \beta \leq 1$; a proof is available at \cite[Lemma~5.2]{Yagan/EG_intersecting_ER}.
On the range $\ell=L_n+1,\ldots,\left \lfloor{\frac{n}{2}}\right \rfloor$, $\left|\nu_{\ell,\ell+1}(\pmb{\alpha})\right|$ can be less than or equal to $L_n$ or greater than $L_n$. 
In the latter case, we have
\begin{equation*}
\max (K_{t_{\min,\ell}},X_{n,\left|\nu_{\ell,\ell+1}(\pmb{\alpha})\right|} +1)\pmb{1}[\left|\nu_{\ell,\ell+1}(\pmb{\alpha})\right| >0]\geq \left \lfloor{\gamma P}\right \rfloor +1
\end{equation*}
Using (\ref{eq:conn_lemma_1}),  (\ref{eq:conn_lemma_1_dash}), and the fact that (see \cite[Lemma~5.4.1]{Yagan/PhD} for a proof)
\begin{equation}\nonumber
{\dbinom{P-K_1}{K_2}}\bigg/{\dbinom{P}{K_2}} \leq e^{-\frac{K_2}{P}K_1}
\label{eq:isolated_BinoExpo}
\end{equation}
for $K_1+K_2 \leq P$,
we have
\begin{align}
&\mathbb{E} \vast[ \frac{\dbinom{P-\max \left(K_{t_{\min,\ell}},X_{n,\left|\nu_{\ell,\ell+1}(\pmb{\alpha})\right|} +1\right)\pmb{1}[\left|\nu_{\ell,\ell+1}(\pmb{\alpha})\right| >0]}{K_j}}{\dbinom {P}{K_j}}  \given[\vast]
\begin{array}{c}
\Sigma_1, \ldots, \Sigma_{\ell},\\
t_{\ell+1}=j
\end{array}\vast] \nonumber \\
&\leq \mathbb{E}\left[ Z_j^{\beta \left|\nu_{\ell,\ell+1}(\pmb{\alpha})\right|} \pmb{1}[\left|\nu_{\ell,\ell+1}(\pmb{\alpha})\right| \leq L_n]  \given[\bigg]
\begin{array}{c}
\Sigma_1, \ldots, \Sigma_{\ell},\\
t_{\ell+1}=j
\end{array}\right] \nonumber \\
& \quad + \mathbb{E} \left[ e^{-\frac{K_j}{P}(\left \lfloor{\gamma P}\right \rfloor +1)} \pmb{1}[\left|\nu_{\ell,\ell+1}(\pmb{\alpha})\right| > L_n]  \given[\bigg]
\begin{array}{c}
\Sigma_1, \ldots, \Sigma_{\ell},\\
t_{\ell+1}=j
\end{array}\right]
\nonumber \\ 
& \leq e^{-\alpha_{\min_j} p_{1j}\beta \ell} + e^{- \gamma K_1} \pmb{1}[\ell>L_n]
\label{eq:conn_lemma}
\end{align}
by virtue of the fact that $K_j \geq K_1$. 

Finally, we note the bounds
\begin{align}
\sum_{j=1}^r \mu_j e^{-\alpha_{\min_j} p_{1j}\beta \ell} &\leq (1-\mu_r)+\mu_r e^{-\alpha_{\min_r} p_{1r} \beta \ell}  \nonumber \\
& \leq (1-\mu_r)+\mu_r e^{-\alpha_{\min} p_{1r} \beta \ell}  \nonumber 
\end{align}
and that
\begin{align}
\sum_{j=1}^r \mu_j e^{-\alpha_{\min_j} p_{1j}\beta \ell} \leq e^{-\alpha_{\min} p_{11} \beta \ell} 
\label{eq:conn_crude_bound_2}
\end{align}
The last step used the fact that $p_{ij}$ is monotone increasing in both $i$ and $j$ and $\alpha_{\min_j} \geq \alpha_{\min}$. 

Note that one could replace $\alpha_{\min}$ with $\alpha_{\min_r}$ in condition (\ref{eq:conn_K1r}) to obtain a more intuitive (and milder) bound that only constraints the product $\min \left\{ \alpha_{1r}(n), \alpha_{2r}(n), \ldots, \alpha_{rr}(n) \right\} p_{1r}(n)$ instead of $\min_{i,j} \left\{ \alpha_{ij}(n) \right\} p_{1r}(n)$. In this case, it would also follow that
\begin{equation}\nonumber
\alpha_{\min_r} p_{1r} \leq \alpha_{mr} p_{mr} = O(\Lambda_m) = O(\log n /n)
\end{equation} 
which is needed in establishing (\ref{eq:conn_key}) along with $\alpha_{\min_r} p_{1r}(n) = \Omega(\log n / n)$ on several ranges of $\ell$ (see Section~\ref{app:establishing}). However, it would still be needed to show that 
\begin{equation}
\sum_{j=1}^r \mu_j e^{-\alpha_{\min_j} p_{1j}\beta \ell} = o(1)
\label{eq:desiredConclusion_modifiedCondition}
\end{equation}
so as to establish (\ref{eq:conn_key}) on the range where $\left \lfloor{\nu n}\right \rfloor+1 \leq \ell \leq \lfloor{\frac{n}{2}} \rfloor $ (see Section~\ref{subsec:range_5}). Observe that on this range, we have
\begin{equation} \nonumber
\sum_{j=1}^r \mu_j e^{-\alpha_{\min_j} p_{1j}\beta \ell}  \leq \sum_{j=1}^r \mu_j e^{- \beta \nu n \alpha_{\min_j} p_{11} } 
\end{equation}
and the desired conclusion (\ref{eq:desiredConclusion_modifiedCondition}) would follow if $n \alpha_{\min_j} p_{11} = \omega(1)$ for $j=1,\ldots,r$. We have (as we invoke (\ref{eq:bounding_variance_of_K}) and the proposed modification of (\ref{eq:conn_K1r}))
\begin{align} 
\alpha_{\min_j} p_{11} &= \frac{\alpha_{\min_j}}{\alpha_{\min_r}} \frac{p_{11}}{p_{1r}} \alpha_{\min_r} p_{1r} = \frac{\alpha_{\min_j}}{\alpha_{\min_r}} \omega \left( \frac{1}{\log n} \right) \Omega \left( \frac{\log n}{n} \right) = \omega \left( \frac{\alpha_{\min_j}}{\alpha_{\min_r}} \frac{1}{n} \right) \nonumber
\end{align}
and the desired conclusion follows if one assumes that
\begin{equation}
\alpha_{\min_j} \sim \alpha_{\min_r}, \quad j=1,\ldots,r-1
\label{eq:eletreby_asEqu_condition}
\end{equation}
i.e., if the per-row minima of the matrix $\pmb{\alpha}$ are all of the same asymptotic order\footnote{This would also give $K_1 = \omega(1)$ since $\alpha_{\min_1} p_{11} = \omega(1/n)$ and $K_1^2 / P = \Omega(p_{11})$. This condition is needed on the range $\left \lfloor{\nu n}\right \rfloor+1 \leq \ell \leq \lfloor{\frac{n}{2}} \rfloor $.}. Indeed, the asymptotic equivalency given in (\ref{eq:eletreby_asEqu_condition}) implies that $\alpha_{\min}(n)$ and $\alpha_{\min_r}(n)$ would need to be on the same order, which essentially translates to (\ref{eq:conn_K1r}). Put differently, establishing (\ref{eq:desiredConclusion_modifiedCondition}) under the modified condition, i.e., $\alpha_{\min_r}(n) p_{1r}(n) = \Omega \left( \log n / n \right)$, requires a new set of asymptotic equivalence conditions that, when combined with the modified condition, are essentially equivalent to (\ref{eq:conn_K1r}).

Next, we establish (\ref{eq:conn_key_bound_aux1}). Let $\mathbb{H}_\ell(n;\pmb{\mu},\pmb{\Theta})$ denote the subgraph of $\mathbb{H}(n;\pmb{\mu},\pmb{\Theta})$ induced on the vertices $\{ v_1,\ldots,v_\ell \}$. $\mathbb{H}_\ell(n;\pmb{\mu},\pmb{\Theta})$ is connected if and only if it contains a spanning tree; i.e., we have
\begin{equation} \nonumber
C_\ell(\pmb{\mu},\pmb{\Theta}) = \cup_{T \in \mathcal{T}_\ell} \left[ T \subseteq \mathbb{H}_\ell(n;\pmb{\mu},\pmb{\Theta}) \right]
\end{equation}
where $\mathcal{T}_\ell$ denotes the collection of all spanning trees on the vertices $\{ v_1,\ldots,v_\ell \}$. 
Thus,
\begin{equation} 
\mathbb{P}[C_\ell(\pmb{\mu},\pmb{\Theta})] \leq \sum_{T \in \mathcal{T}_\ell} \mathbb{P}\left[ T \subseteq \mathbb{H}_\ell(n;\pmb{\mu},\pmb{\Theta}) \right].
\label{eq:app_b_new1}
\end{equation}

Observe that 
\begin{align}
\mathbb{P}\left[ T \subseteq \mathbb{H}_\ell(n;\pmb{\mu},\pmb{\Theta}) \right] &= \mathbb{E}\left[  \mathbb{E}\left[\pmb{1} \left[ T \subseteq \mathbb{H}_\ell(n;\pmb{\mu},\pmb{\Theta})  \right] \given[\big] \Sigma_1, \ldots,\Sigma_\ell \right]   \right] \nonumber \\
& = \mathbb{E} \left[ \mathbb{P}\left[ T \subseteq \mathbb{H}_\ell(n;\pmb{\mu},\pmb{\Theta})  \given[\big] \Sigma_1, \ldots,\Sigma_\ell  \right]    \right] \nonumber \\
& \leq \left( \max_{i,j} \left\{ \alpha_{ij} p_{ij} \right\} \right)^{\ell-1} 
\label{eq:app_b_new2}
\end{align}
where the last inequality follows from the facts that i) a tree on $\ell$ vertices contain $\ell-1$ edges, 
and ii) conditioned on $\Sigma_1, \ldots, \Sigma_\ell$, edge assignments in $\mathbb{H}_\ell(n;\pmb{\mu},\pmb{\Theta})$ are independent and each edge probability is upper bounded by $ \left( \max_{i,j} \left\{ \alpha_{ij} p_{ij} \right\} \right)$. Note that as we use this upper bound, the randomness (stemming from the random variables $\Sigma_1$, $\Sigma_2$, etc.) disappears and (\ref{eq:app_b_new2}) follows. We obtain (\ref{eq:conn_key_bound_aux1}) upon using (\ref{eq:app_b_new2}) in (\ref{eq:app_b_new1}) and noting
by Cayley's formula \cite{Cayley} that
there are $\ell^{\ell-2}$ trees on $\ell$ vertices, i.e.,  $|\mathcal{T}_{\ell}| = \ell^{\ell-2}$. 
\myendpf

\end{document}